\documentclass{article}
\usepackage{amsfonts}

\input{tcilatex}

\begin{document}

\title{\textbf{Orthogonal\ Polynomials\ for\ \ Potentials\ of\ two\
Variables\ \bigskip with External Sources}}
\author{\textbf{M.C.\ BERGERE} \\
Service de Physique Th\'{e}orique, CEA-Saclay\\
F-91191 Gif sur Yvette cedex, France\\
email: bergere@spht.saclay.cea.fr}
\maketitle

\begin{abstract}
This publication is an exercise which extends to two variables the
Christoffel's construction of orthogonal polynomials for potentials of one
variable with external sources. We generalize the construction to
biorthogonal polynomials. We also introduce generalized Schur polynomials as
a set of orthogonal, symmetric, non homogeneous polynomials of several
variables, attached to Young tableaux.
\end{abstract}

\bigskip

\bigskip

\textbf{1. Introduction}

\textbf{\ }

Recent progress in string theory $\left[ 1\right] $ and in quantum
chromodynamics $\left[ 2\right] $ have reactualized the study of the
spectral statistical properties of random matrices. Although most
interesting results are expected from infinitely large matrices, exact
results on the expectation values of ratios of random matrix determinants
have been obtained for finite matrices, using the technique of orthogonal
polynomials.

In this context, considerable results $\left[ 9-10-11\right] $ collected
recently in Ref. $\left[ 3\right] \ $can be found for potentials $V\left(
x\right) \ $of one variable but relatively few and recent results $\left[ 4%
\right] $ are known for potentials of two variables $V\left( x,y\right) $ or
equivalently $V\left( z,z^{\ast }\right) \ $where $z=x+iy;\ $however, such
results could be of some physical interest, for instance, in the BMN program
in string theory $\left[ 5-6\right] .$

It is the purpose of this publication to extend to potentials $V\left(
z,z^{\ast }\right) $ with external sources, the known results for potentials
of one variable. Later on, as it is already done for potentials of one
variable, we expect to generalize this structure to potentials $V\left(
z,z^{\ast }\right) $ with external sources both at the numerator and at the
denominator, using the so called Cauchy transform.

\bigskip

This text is organized as follows: the various results are collected in
Sect. 1, 2 and 3; the proofs are exposed in Sect. 4, 5, 6 and in the
appendix. Finally, a discussion is proposed in Sect. 7 on orthogonal
polynomials for potentials of one and two variables. \bigskip

\bigskip

\bigskip

The construction of orthogonal polynomials for potentials of one variable
with external sources is described as follows: given a real potential \ $%
V\left( x\right) $ which admits an infinite set $\left\{ p_{n}\left(
x\right) \right\} \ $of orthogonal monic polynomials ($n$ is the degree in $%
x\ $of the polynomial) satisfying\qquad 
\begin{equation}
\int dx\ p_{m}\left( x\right) \ p_{n}\left( x\right) \ \ e^{-V\left(
x\right) }=h_{n}\ \ \delta _{nm}  \tag{$1$}
\end{equation}%
we consider $L\ $external sources $\left( \xi _{1},...,\xi _{L}\right) $ and
we look for orthogonal monic polynomials $q_{n}\left( x;\xi _{i}\right) $
such that%
\begin{equation}
\int dx\ q_{m}\left( x;\xi _{i}\right) \ q_{n}\left( x;\xi _{i}\right) \
\dprod\limits_{i=1}^{L}\left( x-\xi _{i}\right) \ \ \ e^{-V\left( x\right)
}=k_{n}\left( \xi _{i}\right) \ \ \delta _{nm}  \tag{$2$}
\end{equation}%
The solution to this problem is known since Christoffel $\left[ 7\right] $
and reads%
\begin{equation}
q_{n}\left( x;\xi _{i}\right) =\frac{1}{\dprod\limits_{i=1}^{L}\left( x-\xi
_{i}\right) }\ \left\vert 
\begin{array}{cccc}
p_{n+L}\left( x\right) & p_{n+L}\left( \xi _{1}\right) & ... & p_{n+L}\left(
\xi _{L}\right) \\ 
... & ... & ... & ... \\ 
p_{n+1}\left( x\right) & p_{n+1}\left( \xi _{1}\right) & ... & p_{n+1}\left(
\xi _{L}\right) \\ 
p_{n}\left( x\right) & p_{n}\left( \xi _{1}\right) & ... & p_{n}\left( \xi
_{L}\right)%
\end{array}%
\right\vert \ \ \frac{1}{\left[ n,L\right] \left( \xi _{i}\right) } 
\tag{$3$}
\end{equation}%
where the determinant%
\begin{equation}
\left[ n,L\right] \left( \xi _{i}\right) =\left\vert 
\begin{array}{ccc}
p_{n+L-1}\left( \xi _{1}\right) & ... & p_{n+L-1}\left( \xi _{L}\right) \\ 
... & ... & ... \\ 
p_{n}\left( \xi _{1}\right) & ... & p_{n}\left( \xi _{L}\right)%
\end{array}%
\right\vert  \tag{$4$}
\end{equation}%
is needed to make the polynomial $q_{n}\left( x;\xi _{i}\right) \ $monic.

\bigskip

The determinant in (3) is a polynomial in $x$ of degree $n+L$ \ which
vanishes for $x=\xi _{i},\ i=1,...,L$;\ therefore, it can be divided by $%
\dprod\limits_{i=1}^{L}\left( x-\xi _{i}\right) $ to give a polynomial in $x$
of degree $n.$ The proof of orthogonality is very simple: we suppose in (2)
that $m<n$ 
\begin{eqnarray}
q_{m}\left( x;\xi _{i}\right) &=&p_{m}\left( x\right)
+\sum_{r=m-1}^{0}a_{r}\left( \xi _{i}\right) \ p_{r}\left( x\right) 
\TCItag{$5$} \\
q_{n}\left( x;\xi _{i}\right) \ \dprod\limits_{i=1}^{L}\left( x-\xi
_{i}\right) &=&p_{n+L}\left( x\right) +\sum_{r=n+L-1}^{n+1}b_{r}\left( \xi
_{i}\right) \ \ p_{r}\left( x\right)  \nonumber \\
&&+\left( -\right) ^{L}\frac{\left[ n+1,L\right] \left( \xi _{i}\right) }{%
\left[ n,L\right] \left( \xi _{i}\right) }\ p_{n}\left( x\right) \  
\TCItag{$6$}
\end{eqnarray}%
The orthogonality of the polynomials $p_{n}\left( x\right) $ ensures the
orthogonality of the polynomials $q_{n}\left( x;\xi _{i}\right) .$

\bigskip

The pseudonorm $k_{n}\left( \xi _{i}\right) $ is found to be the ratio of
two determinants%
\begin{equation}
k_{n}\left( \xi _{i}\right) =\left( -\right) ^{L}\ h_{n}\ \frac{\left[ n+1,L%
\right] \left( \xi _{i}\right) }{\left[ n,L\right] \left( \xi _{i}\right) } 
\tag{$7$}
\end{equation}

\bigskip

\bigskip

\bigskip

Now we state a similar result for real potentials of two variables $x$ and $%
y $.

We introduce the complex variables $z=x+iy,\ z^{\ast }=x-iy,$ and we
consider a real potential $V\left( z,z^{\ast }\right) $ which admits an
infinite set $\left\{ p_{n}\left( z\right) \right\} \ $of orthogonal monic
polynomials%
\begin{equation}
\int \int d^{2}z\ \ p_{m}^{\ast }\left( z\right) \ p_{n}\left( z\right) \
e^{-V\left( z,z^{\ast }\right) }=h_{n}\ \delta _{nm}  \tag{$8$}
\end{equation}%
Everywhere in this text, $\ f^{\ast }\left( z\right) $ is a short notation
for \ $\left[ f\left( z\right) \right] ^{\ast }$. If the domain of
integration and the potential are rotational invariant ($V\left( zz^{\ast
}\right) ),$ the set of orthogonal polynomials are sometimes called
Ginibre's polynomials $\left[ 8\right] $, namely $p_{n}\left( z\right)
=z^{n} $ (of course, the weights $h_{n}$ are different in each case); in
general, the polynomials are not homogeneous and have complex coefficients.

We now consider $L$ external complex sources $\left( \xi _{1},...,\xi
_{L}\right) $ and we look for orthogonal monic polynomials $q_{n}\left(
z;\xi _{i};\xi _{i}^{\ast }\right) $ such that%
\begin{equation}
\int \int d^{2}z\ \ q_{m}^{\ast }\left( z;\xi _{i};\xi _{i}^{\ast }\right) \
q_{n}\left( z;\xi _{i};\xi _{i}^{\ast }\right) \ \ \dprod\limits_{i=1}^{L}\
\left\vert z-\xi _{i}\right\vert ^{2}\ \ \ e^{-V\left( z,z^{\ast }\right)
}=k_{n}\left( \xi _{i};\xi _{i}^{\ast }\right) \ \ \delta _{nm}  \tag{$9$}
\end{equation}%
We show in Sect. 5\ \ that the solution is given by the polynomials%
\begin{eqnarray}
q_{n}\left( z;\xi _{i};\xi _{i}^{\ast }\right) &=&\ \ \left\vert 
\begin{array}{cccc}
Q_{n+L}\left( z;\xi _{i}^{\ast }\right) & Q_{n+L}\left( \xi _{1};\xi
_{i}^{\ast }\right) & ... & Q_{n+L}\left( \xi _{L};\xi _{i}^{\ast }\right)
\\ 
... & ... & ... & ... \\ 
Q_{n+1}\left( z;\xi _{i}^{\ast }\right) & Q_{n+1}\left( \xi _{1};\xi
_{i}^{\ast }\right) & ... & Q_{n+1}\left( \xi _{L};\xi _{i}^{\ast }\right)
\\ 
Q_{n}\left( z;\xi _{i}^{\ast }\right) & Q_{n}\left( \xi _{1};\xi _{i}^{\ast
}\right) & ... & Q_{n}\left( \xi _{L};\xi _{i}^{\ast }\right)%
\end{array}%
\right\vert  \nonumber \\
&&\times \ \dprod\limits_{i=1}^{L}\left( z-\xi _{i}\right) ^{-1}\ \ .\ \ 
\left[ <n,L>\left( \xi _{i};\xi _{i}^{\ast }\right) \right] ^{-1}\  
\TCItag{$10$}
\end{eqnarray}%
where the determinant%
\begin{equation}
<n,L>\left( \xi _{i};\xi _{i}^{\ast }\right) =\left\vert 
\begin{array}{ccc}
Q_{n+L-1}\left( \xi _{1};\xi _{i}^{\ast }\right) & ... & Q_{n+L-1}\left( \xi
_{L};\xi _{i}^{\ast }\right) \\ 
... & ... & ... \\ 
Q_{n}\left( \xi _{1};\xi _{i}^{\ast }\right) & ... & Q_{n}\left( \xi
_{L};\xi _{i}^{\ast }\right)%
\end{array}%
\right\vert  \tag{$11$}
\end{equation}%
is needed to make the polynomial $q_{n}\left( z;\xi _{i};\xi _{i}^{\ast
}\right) $ monic.

\bigskip

The polynomials $Q_{n}\left( z;\xi _{i}^{\ast }\right) $ are monic, of
degree $n$ in $z$ and are defined from the property \ 
\begin{equation}
\int \int d^{2}z\ \ p_{m}^{\ast }\left( z\right) \ Q_{n}\left( z;\xi
_{i}^{\ast }\right) \ \ \dprod\limits_{i=1}^{L}\ \left( z^{\ast }-\xi
_{i}^{\ast }\right) \ \ \ e^{-V\left( z,z^{\ast }\right) }=0\ \ \text{for }%
m<n  \tag{$12$}
\end{equation}%
and of course the corresponding complex conjugate properties. The proof of
the orthogonality \ of the polynomials $q_{n}\left( z;\xi _{i};\xi
_{i}^{\ast }\right) $ is exactly similar to the proof for a potential of one
variable but here, of course, it relies upon the existence of the
polynomials $Q_{n}\left( z;\xi _{i}^{\ast }\right) $ satisfying (12). It is
the purpose of Sect. 4$\ $to show the existence, to construct and to give
the properties of the polynomials $Q_{n}\left( z;\xi _{i}^{\ast }\right) .$
For one external source $\xi $ we show that%
\begin{eqnarray}
\ Q_{n}\left( z;\xi ^{\ast }\right) &=&\frac{h_{n}}{p_{n}^{\ast }\left( \xi
\right) }\ K_{n}\left( z,\xi ^{\ast }\right)  \TCItag{$13$} \\
K_{n}\left( z,\xi ^{\ast }\right) &=&\sum_{i=0}^{n}\frac{1}{h_{i}}\
p_{i}^{\ast }\left( \xi \right) \ p_{i}\left( z\right)  \TCItag{$14$}
\end{eqnarray}%
In the general case with $L$ external sources, we prove that the monic
polynomials%
\begin{equation}
Q_{n}\left( z;\xi _{i}^{\ast }\right) =\frac{h_{n}}{\left[ n,L\right] ^{\ast
}\left( \xi _{i}\right) }\left\vert 
\begin{array}{ccc}
p_{n+L-1}^{\ast }\left( \xi _{1}\right) & ... & p_{n+L-1}^{\ast }\left( \xi
_{L}\right) \\ 
... & ... & ... \\ 
p_{n+1}^{\ast }\left( \xi _{1}\right) & ... & p_{n+1}^{\ast }\left( \xi
_{L}\right) \\ 
K_{n}\left( z,\xi _{1}^{\ast }\right) & ... & K_{n}\left( z,\xi _{L}^{\ast
}\right)%
\end{array}%
\right\vert  \tag{$15$}
\end{equation}%
satisfy (12) and are unique. Clearly, all $K_{n}\left( z,\xi _{i}^{\ast
}\right) $ in (15) could be replaced by the corresponding set $K_{q}\left(
z,\xi _{i}^{\ast }\right) $ for any $q$ satisfying $\ n\leq q<n+L.$

\bigskip

\bigskip We show in Sect. 5 that the polynomials $q_{n}\left( z;\xi _{i};\xi
_{i}^{\ast }\right) $ may be written as 
\begin{eqnarray}
&&q_{n}\left( z;\xi _{i};\xi _{i}^{\ast }\right)  \nonumber \\
&=&\ \left\vert 
\begin{array}{cccc}
p_{n+L}\left( z\right) & p_{n+L}\left( \xi _{1}\right) & ... & p_{n+L}\left(
\xi _{L}\right) \\ 
K_{n+L-1}\left( z,\xi _{1}^{\ast }\right) & K_{n+L-1}\left( \xi _{1},\xi
_{1}^{\ast }\right) & ... & K_{n+L-1}\left( \xi _{L},\xi _{1}^{\ast }\right)
\\ 
... & ... & ... & ... \\ 
K_{n+L-1}\left( z,\xi _{L}^{\ast }\right) & K_{n+L-1}\left( \xi _{1},\xi
_{L}^{\ast }\right) & ... & K_{n+L-1}\left( \xi _{L},\xi _{L}^{\ast }\right)%
\end{array}%
\right\vert  \nonumber \\
&&\times \ \dprod\limits_{i=1}^{L}\left( z-\xi _{i}\right) ^{-1}\ \ .\ \ \ 
\left[ \det \ K_{n+L-1}\left( \xi _{i},\xi _{j}^{\ast }\right) \right]
^{-1}\   \TCItag{$16$}
\end{eqnarray}%
This result has been obtained by Akemann and Vernizzi in Ref. $\left[ 4%
\right] $ (in their work $K_{n}\left( z,\xi ^{\ast }\right) $ is called $%
K_{n+1}\left( z,\overline{\xi }\right) $).

\bigskip

\bigskip \bigskip The norm$\ k_{n}\left( \xi _{i};\xi _{i}^{\ast }\right) $
in (9) is proved to be 
\begin{equation}
k_{n}\left( \xi _{i};\xi _{i}^{\ast }\right) =h_{n+L}\ \frac{\det \
K_{n+L}\left( \xi _{i},\xi _{j}^{\ast }\right) }{\det \ K_{n+L-1}\left( \xi
_{i},\xi _{j}^{\ast }\right) }  \tag{$17$}
\end{equation}%
as conjectured by Akemann and Vernizzi $\left[ 4\right] \ $who verified this
formula up to $L=5.\ $The results of this section are generalized below to
biorthogonal polynomials and all proofs in Sect. 4 and 5 are performed in
that case.

\bigskip

\textbf{2.} \textbf{Biorthogonal Polynomials}

\bigskip

Given two set of external sources $\left( \eta _{1},...,\eta _{L_{1}}\right) 
$ and $\left( \xi _{1},...,\xi _{L_{2}}\right) $, we generalize the above
constructions and define two sets of biorthogonal monic polynomials $%
q_{n}\left( z;\xi _{i};\eta _{i}^{\ast }\right) \ $and $q_{n}\left( z;\eta
_{i};\xi _{i}^{\ast }\right) \ $satisfying the orthogonality relation%
\begin{eqnarray}
&&\int \int d^{2}z\ \ q_{m}^{\ast }\left( z;\eta _{i};\xi _{i}^{\ast
}\right) \ q_{n}\left( z;\xi _{i};\eta _{i}^{\ast }\right) \ \
\dprod\limits_{i=1}^{L_{1}}\left( z^{\ast }-\eta _{i}^{\ast }\right) \
\dprod\limits_{i=1}^{L_{2}}\left( z-\xi _{i}\right) \ \ \ e^{-V\left(
z,z^{\ast }\right) }  \nonumber \\
&=&k_{n}\left( \xi _{i};\eta _{i}^{\ast }\right) \ \ \delta _{nm} 
\TCItag{$18$}
\end{eqnarray}%
For simplicity, we use the same symbol $q_{n}$ for the polynomials $%
q_{n}\left( z;\xi _{i};\eta _{i}^{\ast }\right) \ $and $q_{n}\left( z;\eta
_{i};\xi _{i}^{\ast }\right) $ although they are different functions
depending of the number of variables $\eta _{i}$ and $\xi _{i}.$ The
pseudonorms $k_{n}\left( \xi _{i};\eta _{i}^{\ast }\right) $ are complex and
may vanish on some manifolds of the $\left( \xi _{i};\eta _{i}^{\ast
}\right) $ space. The uniqueness of the biorthogonal polynomials supposes
that we consider generic values of the sources $\xi _{i}$ and$\ \eta
_{i}^{\ast }$ where the norms do not vanish.

\bigskip

The polynomials $q_{n}\left( z;\xi _{i};\eta _{i}^{\ast }\right) $ are
naturally given by%
\begin{eqnarray}
q_{n}\left( z;\xi _{i};\eta _{i}^{\ast }\right) &=&\ \left\vert 
\begin{array}{cccc}
Q_{n+L_{2}}\left( z;\eta _{i}^{\ast }\right) & Q_{n+L_{2}}\left( \xi
_{1};\eta _{i}^{\ast }\right) & ... & Q_{n+L_{2}}\left( \xi _{L_{2}};\eta
_{i}^{\ast }\right) \\ 
... & ... & ... & ... \\ 
Q_{n+1}\left( z;\eta _{i}^{\ast }\right) & Q_{n+1}\left( \xi _{1};\eta
_{i}^{\ast }\right) & ... & Q_{n+1}\left( \xi _{L_{2}};\eta _{i}^{\ast
}\right) \\ 
Q_{n}\left( z;\eta _{i}^{\ast }\right) & Q_{n}\left( \xi _{1};\eta
_{i}^{\ast }\right) & ... & Q_{n}\left( \xi _{L_{2}};\eta _{i}^{\ast }\right)%
\end{array}%
\right\vert \ \   \nonumber \\
&&\times \ \ \dprod\limits_{i=1}^{L_{2}}\left( z-\xi _{i}\right) ^{-1}\ \ .\
\ \left[ <n,L_{2}>\left( \xi _{i};\eta _{i}^{\ast }\right) \right] ^{-1} 
\TCItag{$19$}
\end{eqnarray}%
and correspondingly, the polynomials $q_{n}\left( z;\eta _{i};\xi _{i}^{\ast
}\right) $ are obtained from the polynomials $q_{n}\left( z;\xi _{i};\eta
_{i}^{\ast }\right) $ by exchanging the $\xi $'s and the $\eta $'s, an
operation which implies the exchange of $L_{2}\ $and $L_{1}.$ In (19), $\
<n,L_{2}>\left( \xi _{i};\eta _{i}^{\ast }\right) $ is defined as%
\begin{equation}
<n,L_{2}>\left( \xi _{i};\eta _{i}^{\ast }\right) =\left\vert 
\begin{array}{ccc}
Q_{n+L_{2}-1}\left( \xi _{1};\eta _{i}^{\ast }\right) & ... & 
Q_{n+L_{2}-1}\left( \xi _{L_{2}};\eta _{i}^{\ast }\right) \\ 
... & ... & ... \\ 
Q_{n}\left( \xi _{1};\eta _{i}^{\ast }\right) & ... & Q_{n}\left( \xi
_{L_{2}};\eta _{i}^{\ast }\right)%
\end{array}%
\right\vert  \tag{$20$}
\end{equation}

The polynomials $Q_{n}\left( z;\eta _{i}^{\ast }\right) $ and $Q_{n}\left(
z;\xi _{i}^{\ast }\right) \ $are respectively defined by the following
properties: for $m<n\ $ 
\begin{eqnarray}
\int \int d^{2}z\ \ p_{m}^{\ast }\left( z\right) \ Q_{n}\left( z;\xi
_{i}^{\ast }\right) \ \ \dprod\limits_{i=1}^{L_{2}}\ \left( z^{\ast }-\xi
_{i}^{\ast }\right) \ \ \ e^{-V\left( z,z^{\ast }\right) } &=&0\ \  
\TCItag{$21$} \\
\int \int d^{2}z\ \ p_{m}^{\ast }\left( z\right) \ Q_{n}\left( z;\eta
_{i}^{\ast }\right) \ \ \dprod\limits_{i=1}^{L_{1}}\ \left( z^{\ast }-\eta
_{i}^{\ast }\right) \ \ \ e^{-V\left( z,z^{\ast }\right) } &=&0\ \  
\TCItag{$22$}
\end{eqnarray}%
and the corresponding complex conjugate properties. Again, by simplicity, we
use the same notation $Q_{n}$ in (21) and (22) although they are different
functions depending of the number of variables $\eta _{i}$ and $\xi _{i}$.

The proof of the orthogonality of the polynomials $q$ follows the
corresponding proof for the potentials of one variable; the existence, the
construction and the properties of the polynomials $Q$ are given in Sect.\
4.\ The polynomials $q_{n}\left( z;\xi _{i};\eta _{i}^{\ast }\right) $ take
two different forms depending of the relative values of $L_{1}$ and $L_{2}$;
if $L_{1}\leq L_{2}$ 
\begin{eqnarray}
&&q_{n}\left( z;\xi _{i};\eta _{i}^{\ast }\right)  \nonumber \\
&=&\ \left\vert 
\begin{array}{cccc}
p_{n+L_{2}}\left( z\right) & p_{n+L_{2}}\left( \xi _{1}\right) & ... & 
p_{n+L_{2}}\left( \xi _{L_{2}}\right) \\ 
... & ... & ... & ... \\ 
p_{n+L_{1}}\left( z\right) & p_{n+L_{1}}\left( \xi _{1}\right) & ... & 
p_{n+L_{1}}\left( \xi _{L_{2}}\right) \\ 
K_{n+L_{1}-1}\left( z,\eta _{1}^{\ast }\right) & K_{n+L_{1}-1}\left( \xi
_{1},\eta _{1}^{\ast }\right) & ... & K_{n+L_{1}-1}\left( \xi _{L_{2}},\eta
_{1}^{\ast }\right) \\ 
... & ... & ... & ... \\ 
K_{n+L_{1}-1}\left( z,\eta _{L_{1}}^{\ast }\right) & K_{n+L_{1}-1}\left( \xi
_{1},\eta _{L_{1}}^{\ast }\right) & ... & K_{n+L_{1}-1}\left( \xi
_{L_{2}},\eta _{L_{1}}^{\ast }\right)%
\end{array}%
\right\vert \ \   \nonumber \\
&&\times \ \ \dprod\limits_{i=1}^{L_{2}}\left( z-\xi _{i}\right) ^{-1}\ \ .\
\ \left[ D_{n+L_{1}-1}\left( \xi _{i};\eta _{i}^{\ast }\right) \right] ^{-1}
\TCItag{$23$}
\end{eqnarray}%
and if $L_{1}\geq L_{2}+1$%
\begin{eqnarray}
q_{n}\left( z;\xi _{i};\eta _{i}^{\ast }\right) &=&\frac{h_{n+L_{2}}}{%
\dprod\limits_{i=1}^{L_{2}}\left( z-\xi _{i}\right) }\ \left\vert 
\begin{array}{ccc}
p_{n+L_{1}-1}^{\ast }\left( \eta _{1}\right) & ... & p_{n+L_{1}-1}^{\ast
}\left( \eta _{L_{1}}\right) \\ 
... & ... & ... \\ 
p_{n+L_{2}+1}^{\ast }\left( \eta _{1}\right) & ... & p_{n+L_{2}+1}^{\ast
}\left( \eta _{L_{1}}\right) \\ 
K_{n+L_{2}}\left( z,\eta _{1}^{\ast }\right) & ... & K_{n+L_{2}}\left(
z,\eta _{L_{1}}^{\ast }\right) \\ 
K_{n+L_{2}}\left( \xi _{1},\eta _{1}^{\ast }\right) & ... & 
K_{n+L_{2}}\left( \xi _{1},\eta _{L_{1}}^{\ast }\right) \\ 
... & ... & ... \\ 
K_{n+L_{2}}\left( \xi _{L_{2}},\eta _{1}^{\ast }\right) & ... & 
K_{n+L_{2}}\left( \xi _{L_{2}},\eta _{L_{1}}^{\ast }\right)%
\end{array}%
\right\vert \ \ \   \nonumber \\
&&\times \ \ \frac{1}{D_{n+L_{2}-1}^{\ast }\left( \eta _{i};\xi _{i}^{\ast
}\right) }  \TCItag{$24$}
\end{eqnarray}%
where the determinant $D_{n}\left( \xi _{i};\eta _{i}^{\ast }\right) $ is
defined for $L_{1}\leq L_{2}$ and becomes $D_{n}^{\ast }\left( \eta _{i};\xi
_{i}^{\ast }\right) $ for $L_{1}\geq L_{2}.$ In (23),

\begin{equation}
D_{n}\left( \xi _{i};\eta _{i}^{\ast }\right) =\left\vert 
\begin{array}{ccc}
p_{n+L_{2}-L_{1}}\left( \xi _{1}\right) & ... & p_{n+L_{2}-L_{1}}\left( \xi
_{L_{2}}\right) \\ 
... & ... & ... \\ 
p_{n+1}\left( \xi _{1}\right) & ... & p_{n+1}\left( \xi _{L_{2}}\right) \\ 
K_{n}\left( \xi _{1},\eta _{1}^{\ast }\right) & ... & K_{n}\left( \xi
_{L_{2}},\eta _{1}^{\ast }\right) \\ 
... & ... & ... \\ 
K_{n}\left( \xi _{1},\eta _{L_{1}}^{\ast }\right) & ... & K_{n}\left( \xi
_{L_{2}},\eta _{L_{1}}^{\ast }\right)%
\end{array}%
\right\vert \ \ \ \ \text{for \ \ \ }L_{1}\leq L_{2}  \tag{$25$}
\end{equation}

Clearly, any line of $K_{n}$ can be replaced by the corresponding line of $%
K_{n+q}$ with $0\leq q\leq L_{2}-L_{1}.\ $We show in part 2 of the appendix
that$\ D_{n}\left( \xi _{i};\eta _{i}^{\ast }\right) =0$ for $n<L_{1}-1$ and
that$\ D_{L_{1}-1}\left( \xi _{i};\eta _{i}^{\ast }\right) =$ $%
\dprod\limits_{k=0}^{L_{1}-1}\frac{1}{h_{k}}\ \ \Delta ^{\ast }\left( \eta
_{i}\right) \ \Delta \left( \xi _{i}\right) $ where $\Delta $ is the
Vandermonde determinant defined in (34). Also in (24), 
\begin{equation}
D_{n}^{\ast }\left( \eta _{i};\xi _{i}^{\ast }\right) =\left\vert 
\begin{array}{ccc}
p_{n+L_{1}-L_{2}}^{\ast }\left( \eta _{1}\right) & ... & p_{n+L_{1}-L_{2}}^{%
\ast }\left( \eta _{L_{1}}\right) \\ 
... & ... & ... \\ 
p_{n+1}^{\ast }\left( \eta _{1}\right) & ... & p_{n+1}^{\ast }\left( \eta
_{L_{1}}\right) \\ 
K_{n}\left( \xi _{1},\eta _{1}^{\ast }\right) & ... & K_{n}\left( \xi
_{1},\eta _{L_{1}}^{\ast }\right) \\ 
... & ... & ... \\ 
K_{n}\left( \xi _{L_{2}},\eta _{1}^{\ast }\right) & ... & K_{n}\left( \xi
_{L_{2}},\eta _{L_{1}}^{\ast }\right)%
\end{array}%
\right\vert \ \ \ \ \text{for \ \ \ }L_{1}\geq L_{2}  \tag{$26$}
\end{equation}%
Again, any line of $K_{n}$ can be replaced by the corresponding line of $%
K_{n+q}$ with $0\leq q\leq L_{1}-L_{2}.\ $Similarly, we have $D_{n}^{\ast
}\left( \eta _{i};\xi _{i}^{\ast }\right) =0$ for $n<L_{2}-1$ and $%
D_{L_{2}-1}^{\ast }\left( \eta _{i};\xi _{i}^{\ast }\right) =$ $%
\dprod\limits_{k=0}^{L_{2}-1}\frac{1}{h_{k}}\ \ \Delta ^{\ast }\left( \eta
_{i}\right) \ \Delta \left( \xi _{i}\right) .$

\bigskip The pseudonorm$\ k_{n}\left( \xi _{i};\eta _{i}^{\ast }\right) $ is
found to be

\begin{eqnarray}
k_{n}\left( \xi _{i};\eta _{i}^{\ast }\right) &=&\left( -\right)
^{L_{1}+L_{2}}\ h_{n+L_{1}}\ \frac{D_{n+L_{1}}\left( \xi _{i};\eta
_{i}^{\ast }\right) }{D_{n+L_{1}-1}\left( \xi _{i};\eta _{i}^{\ast }\right) }%
\ \ \ \ \text{for }L_{1}\leq L_{2}  \TCItag{$27$} \\
k_{n}\left( \xi _{i};\eta _{i}^{\ast }\right) &=&\left( -\right)
^{L_{1}+L_{2}}\ h_{n+L_{2}}\ \frac{D_{n+L_{2}}^{\ast }\left( \eta _{i};\xi
_{i}^{\ast }\right) }{D_{n+L_{2}-1}^{\ast }\left( \eta _{i};\xi _{i}^{\ast
}\right) }\ \ \ \ \text{for }L_{1}\geq L_{2}  \TCItag{$28$}
\end{eqnarray}

\bigskip If \bigskip $\ L_{1}=L_{2}\ \ $we recover the result (17) with the $%
\xi _{j}^{\ast }$ replaced by the $\eta _{j}^{\ast }.$\bigskip

\bigskip In Sect. 6 we apply the above formalism to the calculation of the
integrals of the form 
\begin{eqnarray}
&&I_{N,L_{2},L_{1}}\left( \xi _{b};\eta _{a}^{\ast }\right)  \nonumber \\
&=&\int \dprod\limits_{i=1}^{N}d^{2}z_{i}\dprod\limits_{i<j}\left\vert
z_{i}-z_{j}\right\vert ^{2}\dprod\limits_{i,a,b}\left( z_{i}^{\ast }-\eta
_{a}^{\ast }\right) \left( z_{i}-\xi _{b}\right) \ e^{-\sum_{i=1}^{N}V\left(
z_{i},z_{i}^{\ast }\right) }  \TCItag{$29$}
\end{eqnarray}%
where $i$ and $j$ run from $1$ to $N,\ a\ $from $1$ to $L_{1}$ and $b$ from $%
1$ to $L_{2}.\ $These integrals are the generalization to potentials of two
variables of similar integrals with potentials of one variable which have
been expressed by Brezin and Hikami $\left[ 9\right] $ as determinants of
the corresponding orthogonal polynomials. The expressions (31-32) have been
obtained before by Akemann and Vernizzi $\left[ 4\right] .$

It is easily proved, from the property of the Vandermonde determinant (34)
and from the biorthogonality of the polynomials $q_{n}\left( z;\xi _{i};\eta
_{i}^{\ast }\right) ,\ $that%
\begin{equation}
I_{N,L_{2},L_{1}}\left( \xi _{b};\eta _{a}^{\ast }\right) =N!\
\dprod\limits_{n=0}^{N-1}k_{n}\left( \xi _{b};\eta _{a}^{\ast }\right) 
\tag{$30$}
\end{equation}%
This result can be transformed in terms of the determinants$\ D_{n}\left(
\xi _{b};\eta _{a}^{\ast }\right) \ $for \ $L_{1}\leq L_{2}\ $ 
\begin{equation}
I_{N,L_{2},L_{1}}\left( \xi _{b};\eta _{a}^{\ast }\right) =\frac{\left(
-\right) ^{N\left( L_{1}+L_{2}\right) }\ \ N!}{\Delta ^{\ast }\left( \eta
_{a}\right) \ \Delta \left( \xi _{b}\right) }\ \
\dprod\limits_{i=0}^{N+L_{1}-1}h_{i}\ \ \ \ D_{N+L_{1}-1}\left( \xi
_{b};\eta _{a}^{\ast }\right) \ \ \ \ \ \ \   \tag{$31$}
\end{equation}%
and $D_{n}^{\ast }\left( \eta _{a};\xi _{b}^{\ast }\right) $ $\ $for \ $%
L_{1}\geq L_{2}$%
\begin{equation}
I_{N,L_{2},L_{1}}\left( \xi _{b};\eta _{a}^{\ast }\right) =\frac{\left(
-\right) ^{N\left( L_{1}+L_{2}\right) }\ \ N!}{\Delta ^{\ast }\left( \eta
_{a}\right) \ \Delta \left( \xi _{b}\right) }\ \
\dprod\limits_{i=0}^{N+L_{2}-1}h_{i}\ \ \ \ D_{N+L_{2}-1}^{\ast }\left( \eta
_{a};\xi _{b}^{\ast }\right)  \tag{$32$}
\end{equation}

\bigskip

\bigskip

\textbf{3. A Generalization\ of\ the\ Schur\ Polynomials}

We generalize the definition of the Schur polynomials of several variables $%
\left( x_{1},...,x_{n}\right) $ to a set of orthogonal, symmetric and non
homogeneous polynomials of several variables attached to Young tableaux. We
define%
\begin{equation}
P_{\lambda }\left( x_{i}\right) =\frac{\det \ p_{\lambda _{i}+n-i}\left(
x_{j}\right) }{\Delta \left( x_{i}\right) }  \tag{$33$}
\end{equation}%
where $\lambda \ $is a Young tableau described by the length of its rows $%
\lambda _{1}\geq \lambda _{2}\geq ...\geq \lambda _{n}$ and $\Delta \left(
x_{i}\right) $ is the Vandermonde determinant corresponding to the empty
Young tableau%
\begin{equation}
\Delta \left( x_{i}\right) =\dprod\limits_{i<j}\left( x_{i}-x_{j}\right)
=\left\vert 
\begin{array}{ccc}
\pi _{n-1}\left( x_{1}\right) & ... & \pi _{n-1}\left( x_{n}\right) \\ 
... & ... & ... \\ 
\pi _{0}\left( x_{1}\right) & ... & \pi _{0}\left( x_{n}\right)%
\end{array}%
\right\vert  \tag{$34$}
\end{equation}%
for any set of monic polynomials $\left\{ \pi _{n}\left( x\right) \right\} $%
. In the case where the polynomials $p_{n}\left( z\right) \ $are Ginibre$%
^{\prime }$s polynomials, the polynomials $\ P_{\lambda }\left( z_{i}\right) 
$ are Schur polynomials.\ 

The orthogonality of these polynomials is expressed as%
\begin{eqnarray}
&&\int ...\int \dprod\limits_{i=1}^{N}d^{2}z_{i}\
\dprod\limits_{i<j}\left\vert z_{i}-z_{j}\right\vert ^{2}\ P_{\mu }^{\ast
}\left( z_{i}\right) \ \ P_{\nu }\left( z_{i}\right) \
e^{-\sum_{i=1}^{N}V\left( z_{i},z_{i}^{\ast }\right) }  \nonumber \\
&=&N!\ \dprod\limits_{i=1}^{N}h_{\lambda _{i}+N-i\ }\ \ \delta _{\mu \nu } 
\TCItag{$35$}
\end{eqnarray}%
An important relation proved in Sect. 6 is \bigskip 
\begin{equation}
\dprod\limits_{i=1}^{N}\dprod\limits_{a=1}^{L}\left( z_{i}-\xi _{a}\right)
=\sum_{\lambda \in \left[ N\times L\right] }\ \left( -\right)
^{NL-\left\vert \lambda \right\vert }\ \ P_{\lambda }\left( z_{i}\right) \ \
P_{\widetilde{\lambda ^{\prime }}}\left( \xi _{a}\right)  \tag{$36$}
\end{equation}%
where the Young tableau $\lambda ^{\prime }$ is obtained from the Young
tableau $\lambda \ $by exchanging the rows and the columns, and$\ \widetilde{%
\lambda ^{\prime }}$ is the complementary Youg tableau of $\lambda ^{\prime
} $ in the rectangle $\left[ L\times N\right] $;$\ $the surface of the Young
tableau $\lambda $ is denoted by $\left\vert \lambda \right\vert .$

\bigskip

Most of the results of the preceding sections can be expressed in terms of
the generalized Schur polynomials. For instance

\begin{equation}
\left[ n,L\right] \left( \xi _{i}\right) =\Delta \left( \xi _{i}\right) \ \
P_{\left[ L\times n\right] }\left( \xi _{i}\right)  \tag{$37$}
\end{equation}%
where $\left[ L\times n\right] $ is the Young tableau corresponding to the
rectangle with $L$ rows and $n$ columns. Also the polynomials $Q_{n}\left(
z;\eta _{i}^{\ast }\right) $ are found to be%
\begin{equation}
Q_{n}\left( z;\eta _{i}^{\ast }\right) =\frac{h_{n}}{P_{\left[ L\times n%
\right] }^{\ast }\left( \eta _{i}\right) }\ \sum_{j=0}^{n}\ \frac{1}{h_{j}}\
P_{j}\left( z\right) \ \ P_{\left[ L\times n\right] _{j}}^{\ast }\left( \eta
_{i}\right)  \tag{$38$}
\end{equation}%
where the Young tableau $\left[ L\times n\right] _{j}$ is the union of the
rectangle $\left[ L-1\times n\right] $ and of the smallest row $\lambda
_{L}=j.\ $Now, from (62-63) and part 2 of the appendix, we get for $%
L_{1}\leq L_{2}$

\begin{eqnarray}
&<&n,L_{2}>\left( \xi _{i};\eta _{i}^{\ast }\right) =\frac{\Delta \left( \xi
_{i}\right) }{P_{\left[ L_{1}\times n\right] }^{\ast }\left( \eta
_{i}\right) }\dprod\limits_{i=0}^{L_{1}-1}h_{n+i}  \nonumber \\
&&\times \ \sum_{\lambda \in \left[ L_{1}\times n\right] }\dprod%
\limits_{i=1}^{L_{1}}\frac{1}{h_{\lambda _{i}+L_{1}-i}}\ P_{\lambda }^{\ast
}\left( \eta _{i}\right) \ P_{\left[ \left\{ \left( L_{2}-L_{1}\right)
\times n\right\} \cup \lambda \right] }\left( \xi _{i}\right) \ \ \  
\TCItag{$39$}
\end{eqnarray}%
and for $L_{1}\geq L_{2}$ 
\begin{eqnarray}
&<&n,L_{2}>\left( \xi _{i};\eta _{i}^{\ast }\right) =\frac{\Delta \left( \xi
_{i}\right) }{P_{\left[ L_{1}\times n\right] }^{\ast }\left( \eta
_{i}\right) }\dprod\limits_{i=0}^{L_{2}-1}h_{n+i}  \nonumber \\
&&\times \ \sum_{\lambda \in \left[ L_{2}\times n\right] }\dprod%
\limits_{i=1}^{L_{2}}\frac{1}{h_{\lambda _{i}+L_{2}-i}}\ P_{\left[ \left\{
\left( L_{1}-L_{2}\right) \times n\right\} \cup \lambda \right] }^{\ast
}\left( \eta _{i}\right) \ P_{\lambda }\left( \xi _{i}\right) \ \ \  
\TCItag{$40$}
\end{eqnarray}

The polynomials $q_{n}\left( z;\xi _{i};\eta _{i}^{\ast }\right) $ can also
be expressed as the ratio of two such expressions; for $L_{1}\leq L_{2}$ 
\begin{equation}
q_{n}\left( z;\xi _{i};\eta _{i}^{\ast }\right) =\frac{\sum_{\lambda \in %
\left[ L_{1}\times n\right] }\dprod\limits_{i=1}^{L_{1}}\frac{1}{h_{\lambda
_{i}+L_{1}-i}}\ P_{\lambda }^{\ast }\left( \eta _{i}\right) \ P_{\left[
\left\{ \left( L_{2}-L_{1}+1\right) \times n\right\} \cup \lambda \right]
}\left( z,\xi _{i}\right) }{\sum_{\lambda \in \left[ L_{1}\times n\right]
}\dprod\limits_{i=1}^{L_{1}}\frac{1}{h_{\lambda _{i}+L_{1}-i}}\ P_{\lambda
}^{\ast }\left( \eta _{i}\right) \ P_{\left[ \left\{ \left(
L_{2}-L_{1}\right) \times n\right\} \cup \lambda \right] }\left( \xi
_{i}\right) }\ \ \ \ \   \tag{$41$}
\end{equation}

\bigskip and for\ $L_{1}\geq L_{2}+1$%
\begin{eqnarray*}
q_{n}\left( z;\xi _{i};\eta _{i}^{\ast }\right) &=&h_{n+L_{2}\ }\frac{A}{B}
\\
A &=&\sum_{\lambda \in \left[ (L_{2}+1)\times n\right] }\dprod%
\limits_{i=1}^{L_{2}+1}\frac{1}{h_{\lambda _{i}+L_{2}+1-i}}P_{\left[ \left\{
\left( L_{1}-L_{2}-1\right) \times n\right\} \cup \lambda \right] }^{\ast
}\left( \eta _{i}\right) \ P_{\lambda }\left( z,\xi _{i}\right) \\
B &=&\sum_{\lambda \in \left[ L_{2}\times n\right] }\dprod%
\limits_{i=1}^{L_{2}}\frac{1}{h_{\lambda _{i}+L_{2}-i}}\ P_{\left[ \left\{
\left( L_{1}-L_{2}\right) \times n\right\} \cup \lambda \right] }^{\ast
}\left( \eta _{i}\right) \ P_{\lambda }\left( \xi _{i}\right)
\end{eqnarray*}%
\begin{equation}
\tag{$42$}
\end{equation}

In $\left( 41\right) $ and $\left( 42\right) $ the denominators make the
polynomials $q_{n}\left( z;\xi _{i};\eta _{i}^{\ast }\right) $ monic.

Similarly, the pseudonorm $k_{n}\left( \xi _{i};\eta _{i}^{\ast }\right) $
becomes for $L_{1}\leq L_{2}$ 
\begin{eqnarray}
k_{n}\left( \xi _{i};\eta _{i}^{\ast }\right) &=&\left( -\right)
^{L_{1}+L_{2}}\ h_{n+L_{1}}\ \frac{C}{D}  \nonumber \\
C &=&\sum_{\lambda \in \left[ L_{1}\times (n+1)\right] }\dprod%
\limits_{i=1}^{L_{1}}\frac{1}{h_{\lambda _{i}+L_{1}-i}}\ P_{\lambda }^{\ast
}\left( \eta _{i}\right) \ P_{\left[ \left\{ \left( L_{2}-L_{1}\right)
\times (n+1)\right\} \cup \lambda \right] }\left( \xi _{i}\right)  \nonumber
\\
D &=&\sum_{\lambda \in \left[ L_{1}\times n\right] }\dprod%
\limits_{i=1}^{L_{1}}\frac{1}{h_{\lambda _{i}+L_{1}-i}}\ P_{\lambda }^{\ast
}\left( \eta _{i}\right) \ P_{\left[ \left\{ \left( L_{2}-L_{1}\right)
\times n\right\} \cup \lambda \right] }\left( \xi _{i}\right) \ \ \ \ \ \ \ 
\TCItag{$43$}
\end{eqnarray}%
and for\ $L_{1}\geq L_{2}$%
\begin{eqnarray}
k_{n}\left( \xi _{i};\eta _{i}^{\ast }\right) &=&\left( -\right)
^{L_{1}+L_{2}}\ h_{n+L_{2}}\ \frac{E}{F}  \nonumber \\
E &=&\sum_{\lambda \in \left[ L_{2}\times (n+1)\right] }\dprod%
\limits_{i=1}^{L_{2}}\frac{1}{h_{\lambda _{i}+L_{2}-i}}\ P_{\left[ \left\{
\left( L_{1}-L_{2}\right) \times (n+1)\right\} \cup \lambda \right] }^{\ast
}\left( \eta _{i}\right) \ P_{\lambda }\left( \xi _{i}\right)  \nonumber \\
F &=&\sum_{\lambda \in \left[ L_{2}\times n\right] }\dprod%
\limits_{i=1}^{L_{2}}\frac{1}{h_{\lambda _{i}+L_{2}-i}}\ P_{\left[ \left\{
\left( L_{1}-L_{2}\right) \times n\right\} \cup \lambda \right] }^{\ast
}\left( \eta _{i}\right) \ P_{\lambda }\left( \xi _{i}\right)  \TCItag{$44$}
\end{eqnarray}

\bigskip

\bigskip

\bigskip

Finally, we can also express the integrals (29) in terms of the generalized
Schur polynomials; from (31-32) and part 2 of the appendix we obtain for $%
L_{1}\geq L_{2}$%
\begin{eqnarray*}
I_{N,L_{2},L_{1}}\left( \xi _{b};\eta _{a}^{\ast }\right) &=&\left( -\right)
^{N\left( L_{1}+L_{2}\right) }\ N!\ \ \dprod\limits_{i=0}^{N+L_{2}-1}h_{i}\
\  \\
&&\times \sum_{\lambda \in \left[ L_{2}\times N\right] }\dprod%
\limits_{i=1}^{L_{2}}\frac{1}{h_{\lambda _{i}+L_{2}-i}}\ P_{\left[ \left\{
\left( L_{1}-L_{2}\right) \times N\right\} \cup \lambda \right] }^{\ast
}\left( \eta _{a}\right) \ \ P_{\lambda }\left( \xi _{b}\right) \ \ \ 
\end{eqnarray*}%
\begin{equation}
\tag{$45$}
\end{equation}%
and for $L_{1}\leq L_{2}$ 
\begin{eqnarray*}
I_{N,L_{2},L_{1}}\left( \xi _{b};\eta _{a}^{\ast }\right) &=&\left( -\right)
^{N\left( L_{1}+L_{2}\right) }\ N!\ \ \dprod\limits_{i=0}^{N+L_{1}-1}h_{i}\ 
\\
&&\times \ \sum_{\lambda \in \left[ L_{1}\times N\right] }\dprod%
\limits_{i=1}^{L_{1}}\frac{1}{h_{\lambda _{i}+L_{1}-i}}\ P_{\lambda }^{\ast
}\left( \eta _{a}\right) \ \ P_{\left[ \left\{ \left( L_{2}-L_{1}\right)
\times N\right\} \cup \lambda \right] }\left( \xi _{b}\right) \ \ \ 
\end{eqnarray*}%
\begin{equation}
\tag{$46$}
\end{equation}

\bigskip

\bigskip

\bigskip

\textbf{4.}\ \textbf{The\ Polynomials\ }$Q_{n}\left( z;\xi _{i}^{\ast
}\right) \ \ $

\bigskip

\qquad In this section, we look for monic polynomials $Q_{n}\left( z;\xi
_{i}^{\ast }\right) $ satisfying the condition (12). We first show that the
polynomials defined in (15) satisfy the condition (12); then, we prove that
this solution is unique. First, we define the matrix elements%
\begin{equation}
A_{n,m}\left( \xi _{i}^{\ast }\right) =\int \int d^{2}z\ p_{m}^{\ast }\left(
z\right) \ Q_{n}\left( z;\xi _{i}^{\ast }\right) \ e^{-V\left( z,z^{\ast
}\right) }  \tag{$47$}
\end{equation}%
which are zero for $m>n.\ $From (15) we obtain%
\begin{equation}
A_{n,m}\left( \xi _{i}^{\ast }\right) =\frac{h_{n}}{\left[ n,L\right] ^{\ast
}\left( \xi _{i}\right) }\left\vert 
\begin{array}{ccc}
p_{n+L-1}^{\ast }\left( \xi _{1}\right) & ... & p_{n+L-1}^{\ast }\left( \xi
_{L}\right) \\ 
... & ... & ... \\ 
p_{n+1}^{\ast }\left( \xi _{1}\right) & ... & p_{n+1}^{\ast }\left( \xi
_{L}\right) \\ 
p_{m}^{\ast }\left( \xi _{1}\right) & ... & p_{m}^{\ast }\left( \xi
_{L}\right)%
\end{array}%
\right\vert \ \ \ \ \ \text{for\ }m<n+L  \tag{$48$}
\end{equation}%
Clearly, this determinant vanishes for $n<m<n+L.\ $However, it shows that
for $m<n+L\ \ $the elements $A_{n,m}\left( \xi _{i}^{\ast }\right) $ are
linear combinations of the polynomials $p_{m}^{\ast }\left( \xi _{i}\right)
,\ i=1,...,L\ \ $with $L$ coefficients\ independant of $m$.\ 

From the property (34) of the Vandermonde determinant we may write%
\[
p_{m}\left( z\right) \dprod\limits_{i=1}^{L}\left( z-\xi _{i}\right) =\frac{1%
}{\Delta \left( r_{i},\xi _{j}\right) } 
\]%
\begin{equation}
\times \left\vert 
\begin{array}{ccccccc}
p_{m+L}\left( z\right) & p_{m+L}\left( r_{1}\right) & ... & p_{m+L}\left(
r_{m}\right) & p_{m+L}\left( \xi _{1}\right) & ... & p_{m+L}\left( \xi
_{L}\right) \\ 
... & ... & ... & ... & ... & ... & ... \\ 
p_{0}\left( z\right) & p_{0}\left( r_{1}\right) & ... & p_{0}\left(
r_{m}\right) & p_{0}\left( \xi _{1}\right) & ... & p_{0}\left( \xi
_{L}\right)%
\end{array}%
\right\vert  \tag{$49$}
\end{equation}%
where $r_{1},...,r_{m}$ are the roots of the polynomial $p_{m}\left(
z\right) $ and $\Delta \left( r_{i},\xi _{j}\right) $ is the corresponding
Vandermonde determinant$.\ $We now calculate the integral (12) for $m<n;\ $%
we find $\left[ \Delta ^{\ast }\left( r_{i},\xi _{j}\right) \right] ^{-1}$
times the determinant%
\begin{equation}
\left\vert 
\begin{array}{ccccccc}
A_{n,m+L}\left( \xi _{i}^{\ast }\right) & p_{m+L}^{\ast }\left( r_{1}\right)
& ... & p_{m+L}^{\ast }\left( r_{m}\right) & p_{m+L}^{\ast }\left( \xi
_{1}\right) & ... & p_{m+L}^{\ast }\left( \xi _{L}\right) \\ 
... & ... & ... & ... & ... & ... & ... \\ 
A_{n,0}\left( \xi _{i}^{\ast }\right) & p_{0}^{\ast }\left( r_{1}\right) & 
... & p_{0}^{\ast }\left( r_{m}\right) & p_{0}^{\ast }\left( \xi _{1}\right)
& ... & p_{0}^{\ast }\left( \xi _{L}\right)%
\end{array}%
\right\vert  \tag{$50$}
\end{equation}%
This determinant is zero for $m<n\ \ $since, for $q<n+L,\ $the elements $%
A_{n,q}\left( \xi _{i}^{\ast }\right) $ are linear combinations of the
polynomials $p_{q}^{\ast }\left( \xi _{j}\right) ,\ j=1,...,L\ \ $with the
same $q$ independant coefficients$.$\ This achieve the proof of (12) for the
polynomials (15).

\bigskip

\bigskip Before proving the unicity of the polynomials $Q_{n}\left( z;\xi
_{i}^{\ast }\right) ,$ we prove that the integral%
\begin{equation}
\int \int d^{2}z\ \ p_{n}^{\ast }\left( z\right) \ Q_{n}\left( z;\xi
_{i}^{\ast }\right) \ \ \dprod\limits_{i=1}^{L}\ \left( z^{\ast }-\xi
_{i}^{\ast }\right) \ \ \ e^{-V\left( z,z^{\ast }\right) }=\left( -\right)
^{L}\ h_{n}\frac{\left[ n+1,L\right] ^{\ast }\left( \xi _{i}\right) }{\left[
n,L\right] ^{\ast }\left( \xi _{i}\right) }  \tag{$51$}
\end{equation}%
This integral is calculated exactly as above with $m$ replaced by $n$; since 
$L>0$ we obtain$\ \left[ \Delta ^{\ast }\left( r_{i},\xi _{j}\right) \right]
^{-1}$ times the determinant%
\begin{equation}
\left\vert 
\begin{array}{ccccccc}
0 & p_{n+L}^{\ast }\left( r_{1}\right) & ... & p_{n+L}^{\ast }\left(
r_{n}\right) & p_{n+L}^{\ast }\left( \xi _{1}\right) & ... & p_{n+L}^{\ast
}\left( \xi _{L}\right) \\ 
A_{n,n+L-1}\left( \xi _{i}^{\ast }\right) & p_{n+L-1}^{\ast }\left(
r_{1}\right) & ... & p_{n+L-1}^{\ast }\left( r_{n}\right) & p_{n+L-1}^{\ast
}\left( \xi _{1}\right) & ... & p_{n+L-1}^{\ast }\left( \xi _{L}\right) \\ 
... & ... & ... & ... & ... & ... & ... \\ 
A_{n,0}\left( \xi _{i}^{\ast }\right) & p_{0}^{\ast }\left( r_{1}\right) & 
... & p_{0}^{\ast }\left( r_{n}\right) & p_{0}^{\ast }\left( \xi _{1}\right)
& ... & p_{0}^{\ast }\left( \xi _{L}\right)%
\end{array}%
\right\vert  \tag{$52$}
\end{equation}

The same combination of the $p_{q}^{\ast }\left( \xi _{i}\right) $ which
makes the $A_{n,q}\left( \xi _{i}^{\ast }\right) $ for $q=0,...,n+L-1,\ $
transforms (52) into 
\begin{equation}
\left\vert 
\begin{array}{ccccccc}
B_{n,n+L}\left( \xi _{i}^{\ast }\right) & p_{n+L}^{\ast }\left( r_{1}\right)
& ... & p_{n+L}^{\ast }\left( r_{n}\right) & p_{n+L}^{\ast }\left( \xi
_{1}\right) & ... & p_{n+L}^{\ast }\left( \xi _{L}\right) \\ 
0 & p_{n+L-1}^{\ast }\left( r_{1}\right) & ... & p_{n+L-1}^{\ast }\left(
r_{n}\right) & p_{n+L-1}^{\ast }\left( \xi _{1}\right) & ... & 
p_{n+L-1}^{\ast }\left( \xi _{L}\right) \\ 
... & ... & ... & ... & ... & ... & ... \\ 
0 & p_{0}^{\ast }\left( r_{1}\right) & ... & p_{0}^{\ast }\left( r_{n}\right)
& p_{0}^{\ast }\left( \xi _{1}\right) & ... & p_{0}^{\ast }\left( \xi
_{L}\right)%
\end{array}%
\right\vert  \tag{$53$}
\end{equation}%
which is simply equal to$\ \Delta ^{\ast }\left( r_{i},\xi _{j}\right) \
B_{n,n+L}\left( \xi _{i}^{\ast }\right) $ with%
\begin{equation}
B_{n,n+L}\left( \xi _{i}^{\ast }\right) =\left( -\right) ^{L}\ h_{n}\frac{%
\left[ n+1,L\right] ^{\ast }\left( \xi _{i}\right) }{\left[ n,L\right]
^{\ast }\left( \xi _{i}\right) }  \tag{$54$}
\end{equation}

\bigskip This ends the proof of $\left( 51\right) .$

\bigskip

We now prove the uniqueness of the polynomials $Q_{n}\left( z;\xi _{i}^{\ast
}\right) .\ $The proof is a recurrence: $Q_{0}\left( z;\xi _{i}^{\ast
}\right) =1;$ let us suppose that there exists two polynomials $Q_{1}\left(
z;\xi _{i}^{\ast }\right) $ and $Q_{1}^{\prime }\left( z;\xi _{i}^{\ast
}\right) ,$ by difference we obtain%
\begin{equation}
\int \int d^{2}z\ \ \dprod\limits_{i=1}^{L}\ \left( z^{\ast }-\xi _{i}^{\ast
}\right) \ \ \ e^{-V\left( z,z^{\ast }\right) }=0  \tag{$55$}
\end{equation}%
which, from (51), is equivalent to $\left[ 1,L\right] ^{\ast }\left( \xi
_{i}\right) =0.\ $Of course, this is wrong for generic values of the
variables $\xi _{i}.$

We suppose now that we proved that the polynomials $Q_{k}\left( z;\xi
_{i}^{\ast }\right) $ are unique for $k=0,...,n-1$ and consequently given by
(15). If there exists two polynomials $Q_{n}\left( z;\xi _{i}^{\ast }\right) 
$ and $Q_{n}^{\prime }\left( z;\xi _{i}^{\ast }\right) ,$ then by
difference, we have%
\begin{equation}
\int \int d^{2}z\ \ p_{k}^{\ast }\left( z\right) \ Q_{n-1}\left( z;\xi
_{i}^{\ast }\right) \ \ \dprod\limits_{i=1}^{L}\ \left( z^{\ast }-\xi
_{i}^{\ast }\right) \ \ \ e^{-V\left( z,z^{\ast }\right) }=0\ \ \ \ \ \ 
\text{for\ \ }k=0,...,n-1  \tag{$56$}
\end{equation}

Because of the recurrence, the unique polynomial $Q_{n-1}\left( z;\xi
_{i}^{\ast }\right) \ $which satisfy (56) for $k=0,...,n-2$ is given by
(15); Consequently, for $k=n-1$ equation (56) is equivalent to $\left[ n,L%
\right] ^{\ast }\left( \xi _{i}\right) =0\ $which is wrong for generic
values of the variables $\xi _{i}.$

This ends the proof of the uniqueness of the polynomials $Q_{n}\left( z;\xi
_{i}^{\ast }\right) .$

\bigskip

We now establish a first relation for the pseudonorm $k_{n}\left( \xi
_{i};\eta _{i}^{\ast }\right) .$ From the definition (18) we easily obtain
two relations%
\begin{eqnarray}
k_{n}\left( \xi _{i};\eta _{i}^{\ast }\right) &=&\left( -\right) ^{L_{2}}\ 
\frac{<n+1,L_{2}>\left( \xi _{i};\eta _{i}^{\ast }\right) }{<n,L_{2}>\left(
\xi _{i};\eta _{i}^{\ast }\right) }\   \nonumber \\
&&\times \int \int d^{2}z\ \ p_{n}^{\ast }\left( z\right) \ Q_{n}\left(
z;\eta _{i}^{\ast }\right) \ \ \dprod\limits_{i=1}^{L_{1}}\ \left( z^{\ast
}-\eta _{i}^{\ast }\right) \ \ \ e^{-V\left( z,z^{\ast }\right) } 
\TCItag{$57$} \\
k_{n}\left( \xi _{i};\eta _{i}^{\ast }\right) &=&\left( -\right) ^{L_{1}}\ 
\frac{<n+1,L_{1}>^{\ast }\left( \eta _{i};\xi _{i}^{\ast }\right) }{%
<n,L_{1}>^{\ast }\left( \eta _{i};\xi _{i}^{\ast }\right) }\   \nonumber \\
&&\times \int \int d^{2}z\ Q_{n}^{\ast }\left( z;\xi _{i}^{\ast }\right) \
p_{n}\left( z\right) \ \ \dprod\limits_{i=1}^{L_{_{2}}}\ \left( z-\xi
_{i}\right) \ \ \ e^{-V\left( z,z^{\ast }\right) }  \TCItag{$58$}
\end{eqnarray}%
From (51) we may write 
\begin{eqnarray}
k_{n}\left( \xi _{i},\eta _{i}^{\ast }\right) &=&\ \left( -\right)
^{L_{1}+L_{2}}\ \ h_{n}\ \ \frac{\left[ n+1,L_{1}\right] ^{\ast }\left( \eta
_{i}\right) }{\left[ n,L_{1}\right] ^{\ast }\left( \eta _{i}\right) }\  
\nonumber \\
&&\times \frac{<n+1,L_{2}>\left( \xi _{i};\eta _{i}^{\ast }\right) }{%
<n,L_{2}>\left( \xi _{i};\eta _{i}^{\ast }\right) }  \TCItag{$59$} \\
k_{n}\left( \xi _{i},\eta _{i}^{\ast }\right) &=&\ \left( -\right)
^{L_{1}+L_{2}}\ \ h_{n}\ \ \frac{\left[ n+1,L_{2}\right] \left( \xi
_{i}\right) }{\left[ n,L_{2}\right] \left( \xi _{i}\right) }  \nonumber \\
&&\times \ \frac{<n+1,L_{1}>^{\ast }\left( \eta _{i};\xi _{i}^{\ast }\right) 
}{<n,L_{1}>^{\ast }\left( \eta _{i};\xi _{i}^{\ast }\right) }  \TCItag{$60$}
\end{eqnarray}%
These relations will be transformed further later on.

\bigskip

\bigskip

\textbf{5.} \textbf{The\ Polynomials\ \ }$q_{n}\left( z;\xi _{i};\eta
_{i}^{\ast }\right) $

\bigskip

In this section, we calculate the determinants $<n,L_{2}>\left( \xi
_{i};\eta _{i}^{\ast }\right) \ $defined in (20) for the pair of variables $%
\left( \eta _{1},...,\eta _{L_{1}}\right) $ and $\left( \xi _{1},...,\xi
_{L_{2}}\right) $.\ From (15), we see that for $L_{1}\geq 1,\ $the
polynomials $Q_{n}\left( \xi ;\eta _{i}^{\ast }\right) \ $can be developped
as$\ $\ 
\begin{equation}
Q_{n}\left( \xi ;\eta _{i}^{\ast }\right) =\frac{h_{n}}{\left[ n,L_{1}\right]
^{\ast }\left( \eta _{i}\right) }\sum_{j=1}^{L_{1}}\left( -\right)
^{L_{1}-j}\ \left[ n+1,L_{1}-1\right] _{\widehat{j}}^{\ast }\left( \eta
_{i}\right) \ \ \ K_{q}\left( \xi ,\eta _{j}^{\ast }\right)  \tag{$61$}
\end{equation}%
for any $q$ such that $\ n\leq q<n+L_{1}\ $.\ The symbol $\left[ n+1,L_{1}-1%
\right] _{\widehat{j}}^{\ast }\left( \eta _{i}\right) $ means that the
column corresponding to $\eta _{j}$\ is ommited in the determinant; by
convention $\left[ n+1,0\right] _{\widehat{j}}^{\ast }\left( \eta
_{i}\right) =1$. \ 

The rather lengthy calculation of $<n,L_{2}>\left( \xi _{i};\eta _{i}^{\ast
}\right) $ is given in part 1 of the appendix; the expression of the result
depends whether $L_{1}$ is larger or not to $L_{2}.$ We find%
\begin{eqnarray}
&<&n,L_{2}>\left( \xi _{i};\eta _{i}^{\ast }\right) =\frac{%
\dprod\limits_{i=0}^{L_{1}-1}h_{n+i}}{\left[ n,L_{1}\right] ^{\ast }\left(
\eta _{i}\right) }\ D_{n+L_{1}-1}\left( \xi _{i};\eta _{i}^{\ast }\right) 
\text{\ for\ \ }L_{2}\geq L_{1}  \TCItag{$62$} \\
&<&n,L_{2}>\left( \xi _{i};\eta _{i}^{\ast }\right) =\frac{%
\dprod\limits_{i=0}^{L_{2}-1}h_{n+i}}{\left[ n,L_{1}\right] ^{\ast }\left(
\eta _{i}\right) }\ D_{n+L_{2}-1}^{\ast }\left( \eta _{i};\xi _{i}^{\ast
}\right) \text{\ for\ \ }L_{2}\leq L_{1}  \TCItag{$63$}
\end{eqnarray}%
where the determinants $D_{n}\left( \xi _{i};\eta _{i}^{\ast }\right) $ and $%
D_{n}^{\ast }\left( \eta _{i};\xi _{i}^{\ast }\right) $ are defined in
(25-26).

\bigskip

From (19) we may write the polynomials \textbf{\ }$q_{n}\left( z;\xi
_{i};\eta _{i}^{\ast }\right) $ as%
\begin{equation}
q_{n}\left( z;\xi _{i};\eta _{i}^{\ast }\right) =\frac{1}{%
\dprod\limits_{i=1}^{L_{2}}\left( z-\xi _{i}\right) }\ \ \ \frac{%
<n,L_{2}+1>\left( z,\xi _{i};\eta _{i}^{\ast }\right) }{<n,L_{2}>\left( \xi
_{i};\eta _{i}^{\ast }\right) }  \tag{$64$}
\end{equation}%
and from (62-63) we obtain the expressions (23-24). We also get from (59-60
) the pseudonorm $k_{n}\left( \xi _{i},\eta _{i}^{\ast }\right) $ as in
(27-28).

\bigskip

\bigskip

\bigskip

\bigskip

\textbf{6.} \textbf{Applications}

\bigskip

We exploit the existence\ of the biorthogonal polynomials to calculate the
integrals (29).\ We write%
\begin{eqnarray}
\dprod\limits_{i<j}\left\vert z_{i}-z_{j}\right\vert ^{2} &=&\left\vert 
\begin{array}{ccc}
q_{N-1}^{\ast }\left( z_{1};\eta _{i};\xi _{i}^{\ast }\right) & ... & 
q_{N-1}^{\ast }\left( z_{N};\eta _{i};\xi _{i}^{\ast }\right) \\ 
... & ... & ... \\ 
q_{0}^{\ast }\left( z_{1};\eta _{i};\xi _{i}^{\ast }\right) & ... & 
q_{0}^{\ast }\left( z_{N};\eta _{i};\xi _{i}^{\ast }\right)%
\end{array}%
\right\vert  \nonumber \\
&&\times \left\vert 
\begin{array}{ccc}
q_{N-1}\left( z_{1};\xi _{i};\eta _{i}^{\ast }\right) & ... & q_{N-1}\left(
z_{N};\xi _{i};\eta _{i}^{\ast }\right) \\ 
... & ... & ... \\ 
q_{0}\left( z_{1};\xi _{i};\eta _{i}^{\ast }\right) & ... & q_{0}\left(
z_{N};\xi _{i};\eta _{i}^{\ast }\right)%
\end{array}%
\right\vert  \TCItag{$65$}
\end{eqnarray}

Then, we obtain (30) by developping the determinants and by integrating over
each variables $z_{i}\ $using the biorthogonality of the polynomials $q_{n}.$

\bigskip Now, we establish the relation (36) between$\
\dprod\limits_{i=1}^{N}\dprod\limits_{a=1}^{L}\left( z_{i}-\xi _{a}\right) $
and the generalized Schur polynomials (33). First, we write%
\[
\Delta \left( z_{i}\right) \ \Delta \left( \xi _{a}\right)
\dprod\limits_{i=1}^{N}\dprod\limits_{a=1}^{L}\left( z_{i}-\xi _{a}\right) = 
\]%
\begin{equation}
\left\vert 
\begin{array}{cccccc}
p_{N+L-1}\left( z_{1}\right) & ... & p_{N+L-1}\left( z_{N}\right) & 
p_{N+L-1}\left( \xi _{1}\right) & ... & p_{N+L-1}\left( \xi _{L}\right) \\ 
... & ... & ... & ... & ... & ... \\ 
p_{0}\left( z_{1}\right) & ... & p_{0}\left( z_{N}\right) & p_{0}\left( \xi
_{1}\right) & ... & p_{0}\left( \xi _{L}\right)%
\end{array}%
\right\vert  \tag{$66$}
\end{equation}%
Then, we develop the determinant by separating the $z_{i}$ part from the $%
\xi _{a}$ part. We get%
\[
\Delta \left( z_{i}\right) \ \Delta \left( \xi _{a}\right)
\dprod\limits_{i=1}^{N}\dprod\limits_{a=1}^{L}\left( z_{i}-\xi _{a}\right) = 
\]%
\begin{equation}
\sum_{I_{N}}\left( -\right) ^{s\left( I_{n}\right) }\ \left\vert 
\begin{array}{ccc}
p_{i_{N}}\left( z_{1}\right) & ... & p_{i_{N}}\left( z_{N}\right) \\ 
... & ... & ... \\ 
p_{i_{1}}\left( z_{1}\right) & ... & p_{i_{1}}\left( z_{N}\right)%
\end{array}%
\right\vert \ \left\vert 
\begin{array}{ccc}
p_{j_{L}}\left( \xi _{1}\right) & ... & p_{j_{L}}\left( \xi _{L}\right) \\ 
... & ... & ... \\ 
p_{j_{1}}\left( \xi _{1}\right) & ... & p_{j_{1}}\left( \xi _{L}\right)%
\end{array}%
\right\vert  \tag{$67$}
\end{equation}%
where we sum over all subsets of $N\ $indices $I_{N}=\left\{
i_{1}<...<i_{N}\right\} \ $

$\subset \left\{ 0,...,N+L-1\right\} $ and where $\left\{
j_{1}<...<j_{L}\right\} \ $is the complementary subset; the signature for
the sign is%
\begin{equation}
s\left( I_{n}\right) =\frac{N\left( 2L+N-1\right) }{2}-\sum_{a=1}^{N}i_{a} 
\tag{$68$}
\end{equation}%
\ Now, we define two Young tableaux $\lambda $ and $\mu \ $from the length
of their rows%
\begin{eqnarray}
\lambda _{N-a+1} &=&i_{a}-a+1\ \ \ \ \ \ \ \ a=1,...,N  \TCItag{$69$} \\
\mu _{L-b+1} &=&j_{b}-b+1\ \ \ \ \ \ \ \ b=1,...,L  \TCItag{$70$}
\end{eqnarray}%
We observe the relations%
\begin{eqnarray}
\sum_{a=1}^{N}i_{a}+\sum_{b=1}^{L}j_{b} &=&\frac{\left( N+L\right) \left(
N+L-1\right) }{2}  \TCItag{$71$} \\
\left\vert \lambda \right\vert &=&\sum_{a=1}^{N}i_{a}-\frac{N\left(
N-1\right) }{2}  \TCItag{$72$} \\
\left\vert \mu \right\vert &=&\sum_{b=1}^{L}j_{b}-\frac{L\left( L-1\right) }{%
2}  \TCItag{$73$}
\end{eqnarray}%
so that $\left\vert \lambda \right\vert +\left\vert \mu \right\vert =NL.$

Clearly, the Young tableau $\lambda $ belongs to the rectangle $\left[
N\times L\right] $, and the Young tableau $\mu \ $belongs to the rectangle $%
\left[ L\times N\right] .$ Now, we show in part 3 of the appendix that the
complementarity of the indices $\left\{ i_{a}\right\} $ and $\left\{
j_{b}\right\} $ makes $\mu =\widetilde{\lambda ^{\prime }}$ as defined in
Sect. 3 (36). Consequently, (67) is nothing but (36).

We may now calculate $I_{N,L_{2},L_{1}}\left( \xi _{b};\eta _{a}^{\ast
}\right) $ from the orthogonality property (35) of the polynomials $%
P_{\lambda }\left( z_{i}\right) $; we obtain%
\begin{equation}
I_{N,L_{2},L_{1}}\left( \xi _{b};\eta _{a}^{\ast }\right) =\left( -\right)
^{N\left( L_{1}+L_{2}\right) }\ N!\ \sum_{\lambda \in \left[ N\times L\right]
}\dprod\limits_{i=1}^{N}h_{\lambda _{i}+N-i}\ \ P_{\widetilde{\lambda
_{2}^{\prime }}}^{\ast }\left( \xi _{b}\right) \ P_{_{\widetilde{\lambda
_{1}^{\prime }}}}\left( \eta _{a}\right)  \tag{$74$}
\end{equation}%
where$\ \widetilde{\lambda _{1}^{\prime }}$ and $\widetilde{\lambda
_{2}^{\prime }}$ are the complementary Young tableaux of $\lambda ^{\prime
}\ $in the rectangles $\left[ L_{1}\times N\right] $ and $\left[ L_{2}\times
N\right] $ respectively and where $L=Inf\left( L_{1},L_{2}\right) .\ $We
thus obtain (45) or (46) by relabelling $\widetilde{\lambda _{1}^{\prime }}\ 
$or $\widetilde{\lambda _{2}^{\prime }}\ $as $\lambda $ depending whether $%
L_{1}\geq L_{2}$ or not.

Both formulas, (45 or 46) and (74), agree because of the following property
proved in part 3 of the appendix: given a Young tableau $\lambda $ in the
rectangle $\left[ L\times N\right] \ $and the corresponding Young tableau $%
\mu =\ \widetilde{\lambda ^{\prime }}\ $in the rectangle $\left[ N\times L%
\right] ,$ then%
\begin{equation}
\dprod\limits_{i=1}^{L}h_{\lambda _{i}+L-i}\ \ \dprod\limits_{i=1}^{N}h_{\mu
_{i}+N-i}=\dprod\limits_{i=0}^{L+N-1}h_{i}  \tag{$75$}
\end{equation}

\bigskip

\bigskip

\textbf{7. Discussion}

\bigskip

\qquad Given a positive Borel measure $\mu \left( x\right) $ on the real
line and its set of orthogonal, monic polynomials $\left\{ p_{n}\left(
x\right) \right\} $ satisfying%
\begin{equation}
\int \ p_{n}\left( x\right) \ \ p_{m}\left( x\right) \ \ d\mu \left(
x\right) =h_{n}\ \ \delta _{nm}  \tag{$76$}
\end{equation}%
we define two operations:%
\begin{eqnarray}
operation\ 1 &:&d\mu _{1}\left( x\right) =\left( x-\xi \right) \ \ d\mu
\left( x\right)  \TCItag{$77$} \\
operation\ 2 &:&d\mu _{2}\left( x\right) =\ \frac{1}{x-y}\ \ d\mu \left(
x\right)  \TCItag{$78$}
\end{eqnarray}

Although $\mu _{1}\left( x\right) $ and $\mu _{2}\left( x\right) $ are not
positive Borel measures, operations 1 and 2 transform the set of orthogonal
polynomials $p_{n}\left( x\right) \ $into new sets of orthogonal monic
polynomials defined as

- operation 1:%
\begin{eqnarray}
\ p_{n}\left( x\right) &\rightarrow &q_{n}\left( x\right) =\frac{1}{x-\xi }\
\left\vert 
\begin{array}{cc}
p_{n+1}\left( x\right) & p_{n+1}\left( \xi \right) \\ 
p_{n}\left( x\right) & p_{n}\left( \xi \right)%
\end{array}%
\right\vert \ \frac{1}{p_{n}\left( \xi \right) }  \TCItag{$79$} \\
h_{n} &\rightarrow &k_{n}\left( \xi \right) =-\ h_{n}\ \frac{p_{n+1}\left(
\xi \right) }{p_{n}\left( \xi \right) }  \TCItag{$80$}
\end{eqnarray}

\bigskip - operation 2:%
\begin{eqnarray}
p_{n}\left( x\right) &\rightarrow &q_{n}\left( x\right) =\left\vert 
\begin{array}{cc}
p_{n}\left( x\right) & h_{n}\left( y\right) \\ 
p_{n-1}\left( x\right) & h_{n-1}\left( y\right)%
\end{array}%
\right\vert \ \frac{1}{h_{n-1}\left( y\right) }\ \ \ \ \ \ \ \ \ n>0 
\TCItag{$81$} \\
h_{n} &\rightarrow &k_{n}\left( y\right) =-\ h_{n-1}\ \frac{h_{n}\left(
y\right) }{h_{n-1}\left( y\right) }\ \ \ \ \ \ \ n>0  \TCItag{$82$}
\end{eqnarray}%
where $h_{n}\left( y\right) $ is the Cauchy transform of\ $\ p_{n}\left(
x\right) $ defined for $y\notin 
\mathbb{R}
$ as%
\begin{equation}
h_{n}\left( y\right) =\frac{1}{2i\pi }\int \frac{p_{n}\left( x\right) }{x-y}%
\ d\mu \left( x\right)  \tag{$83$}
\end{equation}

\bigskip Of course, $q_{0}\left( x\right) =1$ and $k_{0}\left( y\right)
=2i\pi \ h_{0}\left( y\right) .$

\bigskip

Successive iterations of operation 1 generate Christoffel's results (2-3)
and (7).\ As an application, we obtain Brezin's and Hikami's result $\left[ 9%
\right] $ 
\begin{eqnarray}
&&\int ...\int \dprod\limits_{i=1}^{N}d\mu \left( x_{i}\right) \
\dprod\limits_{i<j}\left( x_{i}-x_{j}\right) ^{2}\
\dprod\limits_{i=1}^{N}\dprod\limits_{a=1}^{L}\left( x_{i}-\xi _{a}\right) 
\nonumber \\
&=&N!\ \dprod\limits_{n=0}^{N-1}k_{n}\left( \xi _{a}\right) =\left( -\right)
^{LN}\ N!\ \dprod\limits_{n=0}^{N-1}h_{n}\ \ \frac{\left[ N,L\right] \left(
\xi _{a}\right) }{\Delta \left( \xi _{a}\right) }  \TCItag{$84$}
\end{eqnarray}%
Successive iterations of operations 1 and 2 generate a set of orthogonal
monic polynomials described by Uvarov $\left[ 10\right] ,\ $namely, for a
given set of external sources $\left( \xi _{1},...,\xi _{L}\right) \ $at the
numerator and $\left( y_{1},...,y_{M}\right) $ at the denominator, we have%
\begin{equation}
\int d\mu \left( x\right) \ q_{n}\left( x\right) \ q_{m}\left( x\right) \ 
\frac{\dprod\limits_{i=1}^{L}\left( x-\xi _{i}\right) }{\dprod%
\limits_{j=1}^{M}\left( x-y_{j}\right) }=k_{n}\left( \xi _{i};y_{j}\right) \
\delta _{nm}  \tag{$85$}
\end{equation}%
The orthogonal polynomials are, for $n\geq M$%
\[
q_{n}\left( x\right) =\dprod\limits_{i=1}^{L}\left( x-\xi _{i}\right) ^{-1}\
\ \left[ \left[ n-M,L+M\right] \left( \xi _{i};y_{j}\right) \right] ^{-1}\
\times 
\]%
\begin{equation}
\left\vert 
\begin{array}{ccccccc}
p_{n+L}\left( x\right) & p_{n+L}\left( \xi _{1}\right) & ... & p_{n+L}\left(
\xi _{L}\right) & h_{n+L}\left( y_{1}\right) & ... & h_{n+L}\left(
y_{M}\right) \\ 
... & ... & ... & ... & ... & ... & ... \\ 
p_{n-M}\left( x\right) & p_{n-M}\left( \xi _{1}\right) & ... & p_{n-M}\left(
\xi _{L}\right) & h_{n-M}\left( y_{1}\right) & ... & h_{n-M}\left(
y_{M}\right)%
\end{array}%
\right\vert  \tag{$86$}
\end{equation}%
and%
\begin{equation}
k_{n}\left( \xi _{i};y_{j}\right) =\left( -\right) ^{L+M}\ h_{n-M}\ \frac{%
\left[ n-M+1,L+M\right] \left( \xi _{i};y_{j}\right) }{\left[ n-M,L+M\right]
\left( \xi _{i};y_{j}\right) }  \tag{$87$}
\end{equation}
where $\left[ n-M,L+M\right] \left( \xi _{i};y_{j}\right) $ is the minor of $%
p_{n+L}\left( x\right) $ in the determinant (86). For $n<M$ \ equations
(86-87) are replaced by%
\[
q_{n}\left( x\right) =\dprod\limits_{i=1}^{L}\left( x-\xi _{i}\right) ^{-1}\
\ \left[ \left[ n-M,L+M\right] \left( \xi _{i};y_{j}\right) \right] ^{-1}\
\times 
\]%
\begin{equation}
\left\vert 
\begin{array}{ccccccc}
p_{n+L}\left( x\right) & p_{n+L}\left( \xi _{1}\right) & ... & p_{n+L}\left(
\xi _{L}\right) & h_{n+L}\left( y_{1}\right) & ... & h_{n+L}\left(
y_{M}\right) \\ 
... & ... & ... & ... & ... & ... & ... \\ 
p_{0}\left( x\right) & p_{0}\left( \xi _{1}\right) & ... & p_{0}\left( \xi
_{L}\right) & h_{0}\left( y_{1}\right) & ... & h_{0}\left( y_{M}\right) \\ 
0 & 0 & ... & 0 & p_{M-n-1}\left( y_{1}\right) & ... & p_{M-n-1}\left(
y_{M}\right) \\ 
... & ... & ... & ... & ... & ... & ... \\ 
0 & 0 & ... & 0 & p_{0}\left( y_{1}\right) & ... & p_{0}\left( y_{M}\right)%
\end{array}%
\right\vert  \tag{$88$}
\end{equation}%
and%
\begin{equation}
k_{n}\left( \xi _{i};y_{j}\right) =\left( -\right) ^{L+n}\ 2i\pi \ \ \frac{%
\left[ n-M+1,L+M\right] \left( \xi _{i};y_{j}\right) }{\left[ n-M,L+M\right]
\left( \xi _{i};y_{j}\right) }  \tag{$89$}
\end{equation}%
We find convenient to keep the notation $\left[ n-M,L+M\right] \left( \xi
_{i};y_{j}\right) $ for the minor of $p_{n+L}\left( x\right) $ in the
determinant (88) although $\left( n-M\right) <0;\ $in this notation, $\left(
M-n\right) \ $is the number of lines of zeroes and of $p_{q}\left(
y_{a}\right) $\ and $\left( L+M\right) $ is the size of the\ determinant. As 
$n$ increases to $M$ the number of lines of zeroes and of $p_{q}\left(
y_{a}\right) $ decreases and that provides a natural transition from (88) to
(86) as $n\ $becomes larger than $M.$ As a special case, we have $\left[
-M,L+M\right] \left( \xi _{i};y_{j}\right) =\Delta \left( \xi _{i}\right) \
\Delta \left( y_{j}\right) .$

We note that in Uvarov's expression for (88), $p_{q}\left( y_{a}\right) $
for $0\leq q\leq M-n-1$ is replaced by $y_{a}^{q}\ $(which is the same
situation as in the Vandermonde determinant and leaves the value of the
determinant unchanged).

\bigskip

\qquad As a consequence of this set of orthogonal polynomials, we can
generalize the calculation of the integrals (84) to the integrals%
\begin{equation}
I_{N}=\int ...\int \dprod\limits_{i=1}^{N}d\mu \left( x_{i}\right) \
\dprod\limits_{i<j}\left( x_{i}-x_{j}\right) ^{2}\ \frac{\dprod%
\limits_{i=1}^{N}\dprod\limits_{a=1}^{L}\left( x_{i}-\xi _{a}\right) }{%
\dprod\limits_{i=1}^{N}\dprod\limits_{b=1}^{M}\left( x_{i}-y_{b}\right) }%
=N!\ \dprod\limits_{n=0}^{N-1}k_{n}\left( \xi _{a};y_{b}\right)  \tag{$90$}
\end{equation}

\bigskip For $N>M$, equation (90) is%
\begin{equation}
I_{N}=N!\ \left( -\right) ^{\left( L+M\right) N}\ \left( -\right) ^{\frac{%
M\left( M+1\right) }{2}}\ \left( 2i\pi \right) ^{M}\ \frac{%
\dprod\limits_{i=0}^{N-M-1}h_{i}}{\Delta \left( \xi _{a}\right) \ \Delta
\left( y_{b}\right) }\ \left[ N-M,L+M\right] \left( \xi _{i};y_{j}\right) 
\tag{$91$}
\end{equation}%
this result has been obtained by Fyodorov and Strahov$\ \left[ 11\right] ;$
for $N\leq M$, we obtain%
\begin{equation}
I_{N}=N!\ \left( -\right) ^{LN}\ \left( -\right) ^{\frac{N\left( N-1\right) 
}{2}}\ \left( 2i\pi \right) ^{N}\ \frac{1}{\Delta \left( \xi _{a}\right) \
\Delta \left( y_{b}\right) }\ \left[ N-M,L+M\right] \left( \xi
_{i};y_{j}\right)  \tag{$92$}
\end{equation}

\bigskip

Of course, when some $\xi $'s and some $y$'s are equal, equations (91-92)
provide many relations between determinants. Moreover, if $L=M,$ it is shown
in the appendix (part 4) that equations (91-92) can be expressed in terms of
the determinant of a$\ \left[ M\times M\right] $ matrix; for$\ N\geq M$, 
\begin{eqnarray}
I_{N} &=&N!\ \ \left( -\right) ^{\frac{M\left( M-1\right) }{2}}\left(
\dprod\limits_{i=0}^{N-1}h_{i}\ \right) \ \frac{\dprod\limits_{a=1}^{M}%
\dprod\limits_{b=1}^{M}\left( y_{b}-\xi _{a}\right) }{\Delta \left( \xi
_{a}\right) \ \Delta \left( y_{b}\right) }\   \nonumber \\
&&\times \ \det \ \left[ 2i\pi \ H_{N-1}\left( \xi _{a};y_{b}\right) +\frac{1%
}{y_{b}-\xi _{a}}\right]  \TCItag{$93$}
\end{eqnarray}%
where 
\begin{equation}
H_{N}\left( \xi _{a};y_{b}\right) =\sum_{i=0}^{N}\frac{1}{h_{i}}\
p_{i}\left( \xi _{a}\right) \ h_{i}\left( y_{b}\right)  \tag{$94$}
\end{equation}

\bigskip

Similarly, for $N\leq M$ \ it is shown in part 4 of the appendix that%
\begin{equation}
I_{N}=N!\ \ \left( -\right) ^{\frac{N\left( N-1\right) }{2}}\ \frac{%
\dprod\limits_{i=0}^{N-1}h_{i}\ }{\Delta \left( \theta _{a}\right) \ \Delta
\left( y_{b}\right) }\ \ \left\vert 
\begin{array}{ccc}
U_{N-1}\left( \theta _{1},y_{1}\right) & ... & U_{N-1}\left( \theta
_{1},y_{M}\right) \\ 
... & ... & ... \\ 
U_{N-1}\left( \theta _{N},y_{1}\right) & ... & U_{N-1}\left( \theta
_{N},y_{M}\right) \\ 
p_{M-N-1}\left( y_{1}\right) & ... & p_{M-N-1}\left( y_{M}\right) \\ 
... & ... & ... \\ 
p_{0}\left( y_{1}\right) & ... & p_{0}\left( y_{M}\right)%
\end{array}%
\right\vert  \tag{$95$}
\end{equation}%
where%
\begin{equation}
U_{N-1}\left( \theta _{a},y_{j}\right) =\dprod\limits_{i=1}^{M}\left(
y_{j}-\xi _{i}\right) \ \ \left[ 2i\pi \ H_{N-1}\left( \theta
_{a};y_{j}\right) +\frac{1}{y_{j}-\theta _{a}}\right]  \tag{$96$}
\end{equation}%
and where the set of $N$ variables $\theta _{a}$ is any subset of $\left(
\xi _{1},...,\xi _{M}\right) .$If $M=N$ \ \ there is no more\ $p_{q}\left(
y_{j}\right) $ part in (95) and we recover(93).

Now, if we take in (90) and (93) the derivatives of $I_{N}$\ in regards to
all variables $\xi _{a},$ then, if we make all variables $\xi _{a}=y_{a},$
we prove in part 4 of the appendix that for $N\geq M$ 
\begin{eqnarray}
\bigskip J_{N} &=&\int ...\int \dprod\limits_{i=1}^{N}d\mu \left(
x_{i}\right) \ \dprod\limits_{i<j}\left( x_{i}-x_{j}\right) ^{2}\
\dprod\limits_{a=1}^{M}\left( \sum_{i=1}^{N}\frac{1}{x_{i}-\xi _{a}}\right) 
\nonumber \\
&=&N!\ \ \dprod\limits_{i=0}^{N-1}h_{i}\ \ \ "\det "T_{M}\left( \xi _{a};\xi
_{b}\right)  \TCItag{$97$}
\end{eqnarray}%
where$\ T_{M}\left( \xi _{a};\xi _{b}\right) $ is a $\left[ M\times M\right] 
$ matrix, the elements of which are%
\begin{eqnarray}
\left[ T_{M}\right] _{aa} &=&2i\pi \ H_{N-1}\left( \xi _{a};\xi _{a}\right) 
\TCItag{$98$} \\
\left[ T_{M}\right] _{a\neq b} &=&2i\pi \ H_{N-1}\left( \xi _{a};\xi
_{b}\right) +\frac{1}{\xi _{b}-\xi _{a}}  \TCItag{$99$}
\end{eqnarray}%
and where $"\det "$ is a notation which means that we ignore the double
poles at $\xi _{a}=\xi _{b}\ $as we develop the determinant. Moreover, it is
easy to see that the residues of the single poles at $\xi _{a}=\xi _{b}$ are
zero, so that the integrals (97) are analytic in $\xi $'s if the $\xi $'s $%
\notin 
\mathbb{R}
.$

$\bigskip $

The discontinuity of $h_{n}\left( y\right) $ when $y$ $\in 
\mathbb{R}
$ is given by (83)%
\begin{equation}
disc\ h_{n}\left( y\right) =p_{n}\left( y\right) \ \mu ^{\prime }\left(
y\right)  \tag{$100$}
\end{equation}%
where we use the notation $d\mu \left( y\right) =$ $\mu ^{\prime }\left(
y\right) \ dy\ .$Consequently, for $N\geq M\ \ $the total discontinuity of $%
J_{N}$ , for a set of different real variables $\xi ^{\prime }$s (which is
related to the $N$-point correlation function), is%
\begin{equation}
disc_{\xi _{1},...,\xi _{M}}\ J_{N}=N!\ \ \left(
\dprod\limits_{i=0}^{N-1}h_{i}\ \right) \ \left( 2i\pi \right) ^{M}\ \
\left( \dprod\limits_{i=1}^{M}\mu ^{\prime }\left( \xi _{i}\right) \right)
\det \ K_{N-1}\left( \xi _{a};\xi _{b}\right)  \tag{$101$}
\end{equation}%
as already obtained in Ref. $\left[ 12-13\right] .\ $For $N<M$ \ the
expression for $J_{N}$ is more complicate but the total discontinuity in
that case is clearly zero.

\bigskip

\bigskip\ 

\bigskip

\qquad A similar program should be achieved for potentials of two variables $%
V\left( z,z^{\ast }\right) $.

We consider a positive Borel measure $\mu \left( z,z^{\ast }\right) \ $on
the complex plane and its set of orthogonal, monic polynomials $\left\{
p_{n}\left( z\right) \right\} $ satisfying%
\begin{equation}
\int \ p_{n}^{\ast }\left( z\right) \ \ p_{m}\left( z\right) \ \ d\mu \left(
z,z^{\ast }\right) =\ h_{n}\ \ \delta _{nm}  \tag{$102$}
\end{equation}%
We define four operations:%
\begin{eqnarray}
operation1 &:&d\mu _{1}\left( z,z^{\ast }\right) =\left( z-\xi \right) \
d\mu \left( z,z^{\ast }\right)  \TCItag{$103$} \\
operation2 &:&d\mu _{2}\left( z,z^{\ast }\right) =\left( z^{\ast }-\eta
^{\ast }\right) \ d\mu \left( z,z^{\ast }\right)  \TCItag{$104$} \\
operation3 &:&d\mu _{3}\left( z,z^{\ast }\right) =\frac{1}{\left( z-y\right) 
}\ d\mu \left( z,z^{\ast }\right)  \TCItag{$105$} \\
operation\ 4 &:&d\mu _{4}\left( z,z^{\ast }\right) =\frac{1}{\left( z^{\ast
}-x^{\ast }\right) }\ d\mu \left( z,z^{\ast }\right)  \TCItag{$106$}
\end{eqnarray}

Operations 1 and 2 transform the set of orthogonal polynomials $p_{n}\left(
z\right) $ into sets of biorthogonal monic polynomials defined as

- operation 1:%
\begin{eqnarray}
p_{n}\left( z\right) &\rightarrow &q_{n}\left( z;\xi ;\Phi \right) =\frac{1}{%
\left( z-\xi \right) }\ \left\vert 
\begin{array}{cc}
p_{n+1}\left( z\right) & p_{n+1}\left( \xi \right) \\ 
p_{n}\left( z\right) & p_{n}\left( \xi \right)%
\end{array}%
\right\vert \ \frac{1}{p_{n}\left( \xi \right) }  \TCItag{$107$} \\
p_{n}^{\ast }\left( z\right) &\rightarrow &q_{n}^{\ast }\left( z;\Phi ;\xi
^{\ast }\right) =Q_{n}^{\ast }\left( z;\xi ^{\ast }\right) =\frac{h_{n}}{%
p_{n}\left( \xi \right) }\ K_{n}^{\ast }\left( z,\xi ^{\ast }\right) 
\TCItag{$108$} \\
h_{n} &\rightarrow &k_{n}\left( \xi \right) =-h_{n}\ \frac{p_{n+1}\left( \xi
\right) }{p_{n}\left( \xi \right) }  \TCItag{$109$}
\end{eqnarray}%
In (107-108), $\Phi $ means an empty set of variables, $Q_{n}$ and $K_{n}\ $%
are defined in (13-14) and have the property that%
\begin{equation}
\int Q_{n}^{\ast }\left( z;\xi ^{\ast }\right) \ \ p_{m}\left( z\right) \ \
\left( z-\xi \right) \ \ d\mu \left( z,z^{\ast }\right) =0\ \ \ \text{for}\
\ m<n  \tag{$110$}
\end{equation}

\bigskip

- operation 2:\bigskip 
\begin{eqnarray}
p_{n}\left( z\right) &\rightarrow &q_{n}\left( z;\Phi ;\eta ^{\ast }\right)
=Q_{n}\left( z;\eta ^{\ast }\right) =\frac{h_{n}}{p_{n}^{\ast }\left( \eta
\right) }\ K_{n}\left( z,\eta ^{\ast }\right)  \TCItag{$111$} \\
p_{n}^{\ast }\left( z\right) &\rightarrow &q_{n}^{\ast }\left( z;\eta ;\Phi
\right) =\frac{1}{\left( z^{\ast }-\eta ^{\ast }\right) }\ \left\vert 
\begin{array}{cc}
p_{n+1}^{\ast }\left( z\right) & p_{n+1}^{\ast }\left( \eta \right) \\ 
p_{n}^{\ast }\left( z\right) & p_{n}^{\ast }\left( \eta \right)%
\end{array}%
\right\vert \ \frac{1}{p_{n}^{\ast }\left( \eta \right) }  \TCItag{$112$} \\
h_{n} &\rightarrow &k_{n}\left( \eta ^{\ast }\right) =-h_{n}\ \frac{%
p_{n+1}^{\ast }\left( \eta \right) }{p_{n}^{\ast }\left( \eta \right) } 
\TCItag{$113$}
\end{eqnarray}%
where, similarly to (110), we have%
\begin{equation}
\int \ p_{m}^{\ast }\left( z\right) \ \ Q_{n}\left( z;\eta ^{\ast }\right) \
\ \left( z^{\ast }-\eta ^{\ast }\right) \ \ d\mu \left( z,z^{\ast }\right)
=0\ \ \ \text{for}\ \ m<n  \tag{$114$}
\end{equation}

\bigskip

Successive iterations of operations 1 and 2 is the purpose of this
publication and generate the biorthogonal polynomials $q_{n}\left( z;\xi
_{i};\eta _{i}^{\ast }\right) $\ and $q_{n}^{\ast }\left( z;\eta _{i};\xi
_{i}^{\ast }\right) $ described in (18-19). Operations 3 and 4 are beyond
the scope of this publication and might be described in a near future.

\bigskip

A\textit{cknowledgment.} I thank G. Akemann and B. Eynard for useful
discussions and for a careful reading of the manuscript. B.\ Eynard
encourages me to extend the work to biorthogonal polynomials. I also thank
D. Giuliano who is at the origin of my interest to the subject.

\bigskip

\bigskip

\bigskip

\bigskip

\bigskip

\textbf{Appendix}

\bigskip

\bigskip

\textsl{Part\ 1:}\textit{\ }Given two sets of source variables $\left( \eta
_{1},...,\eta _{L_{1}}\right) $ and $\left( \xi _{1},...,\xi _{L_{2}}\right)
,\ $this part of the appendix is devoted to the calculation of $%
<n,L_{2}>\left( \xi _{i};\eta _{i}^{\ast }\right) $ \ defined in (20).

1$%
{{}^\circ}%
)\ $We$\ $first suppose that $\ L_{2}\geq L_{1}.\ $From (20) and (61) we
write%
\begin{eqnarray}
&<&n,L_{2}>\left( \xi _{i};\eta _{i}^{\ast }\right)
=\dprod\limits_{i=0}^{L_{2}-1}\frac{h_{n+i}}{\left[ n+i,L_{1}\right] ^{\ast
}\left( \eta _{i}\right) }  \nonumber \\
&&\times \sum_{\left\{ j_{k}\right\} =1}^{L_{1}}\left( -\right)
^{L_{1}L_{2}-\sum j_{k}}\dprod\limits_{k=1}^{L_{2}}\left[ n+1+L_{2}-k,L_{1}-1%
\right] _{\widehat{j_{k}}}^{\ast }\left( \eta _{i}\right)  \nonumber \\
&&\times \ \left\vert 
\begin{array}{ccc}
K_{n+L_{2}-1}\left( \xi _{1},\eta _{j_{1}}^{\ast }\right) & ... & 
K_{n+L_{2}-1}\left( \xi _{L_{2}},\eta _{j_{1}}^{\ast }\right) \\ 
... & ... & ... \\ 
K_{n+L_{1}}\left( \xi _{1},\eta _{j_{L_{2}-L_{1}}}^{\ast }\right) & ... & 
K_{n+L_{1}}\left( \xi _{L_{2}},\eta _{j_{L_{2}-L_{1}}}^{\ast }\right) \\ 
K_{n+L_{1}-1}\left( \xi _{1},\eta _{j_{L_{2}-L_{1}+1}}^{\ast }\right) & ...
& K_{n+L_{1}-1}\left( \xi _{L_{2}},\eta _{j_{L_{2}-L_{1}+1}}^{\ast }\right)
\\ 
... & ... & ... \\ 
K_{n+L_{1}-1}\left( \xi _{1},\eta _{j_{L_{2}}}^{\ast }\right) & ... & 
K_{n+L_{1}-1}\left( \xi _{L_{2}},\eta _{j_{L_{2}}}^{\ast }\right)%
\end{array}%
\right\vert  \TCItag{$A1$}
\end{eqnarray}%
where we sum over all indices $\left\{ j_{1},...,j_{L_{2}}\right\} $ running
from $1$ to $L_{1}.$ In the\ $L_{1}\ $bottom lines of the determinant, the $%
\eta _{i}^{\ast }\ $variables in the functions $K_{n+L_{1}-1}$ take all
possible values $\eta _{1}^{\ast },...,\eta _{L_{1}}^{\ast }.$\
Consequently, $K_{n+L_{1}}\left( \xi _{i},\eta _{j_{L_{2}-L_{1}}}^{\ast
}\right) $ can be replaced by $\frac{p_{n+L_{1}}\left( \xi _{i}\right) \
p_{n+L_{1}}^{\ast }\left( \eta _{j_{L_{2}-L_{1}}}\right) }{h_{n+L_{1}}};$
furthemore, $K_{n+L_{1}+1}\left( \xi _{i},\eta _{j_{L_{2}-L_{1}-1}}^{\ast
}\right) $ can be replaced by $\frac{p_{n+L_{1}+1}\left( \xi _{i}\right) \
p_{n+L_{1}+1}^{\ast }\left( \eta _{j_{L_{2}-L_{1}-1}}\right) }{h_{n+L_{1}-1}}%
\ $\ since the terms $p_{n+L_{1}}\left( \xi _{i}\right) $ are now
proportional in the lines $n+L_{1}$ and $n+L_{1}+1.\ $The argument follows
up to $N+L_{2}-1;$\ \ we may now write%
\begin{eqnarray}
&<&n,L_{2}>\left( \xi _{i};\eta _{i}^{\ast }\right)
=\dprod\limits_{i=0}^{L_{2}-1}\frac{h_{n+i}}{\left[ n+i,L_{1}\right] ^{\ast
}\left( \eta _{i}\right) }\sum_{\left\{ j_{k}\right\} =1}^{L_{1}}\left(
-\right) ^{L_{1}L_{2}-\sum j_{k}}  \nonumber \\
&&\dprod\limits_{k=1}^{L_{2}}\left[ n+1+L_{2}-k,L_{1}-1\right] _{\widehat{%
j_{k}}}^{\ast }\left( \eta _{i}\right) \dprod\limits_{k=1}^{L_{2}-L_{1}}%
\frac{p_{n+L_{2}-k}^{\ast }\left( \eta _{j_{k}}\right) }{h_{n+L_{2}-k}} 
\nonumber \\
&&\ \ \ \left\vert 
\begin{array}{ccc}
p_{n+L_{2}-1}\left( \xi _{1}\right) & ... & p_{n+L_{2}-1}\left( \xi
_{L_{2}}\right) \\ 
... & ... & ... \\ 
p_{n+L_{1}}\left( \xi _{1}\right) & ... & p_{n+L_{1}}\left( \xi
_{L_{2}}\right) \\ 
K_{n+L_{1}-1}\left( \xi _{1},\eta _{j_{L_{2}-L_{1}+1}}^{\ast }\right) & ...
& K_{n+L_{1}-1}\left( \xi _{L_{2}},\eta _{j_{L_{2}-L_{1}+1}}^{\ast }\right)
\\ 
... & ... & ... \\ 
K_{n+L_{1}-1}\left( \xi _{1},\eta _{j_{L_{2}}}^{\ast }\right) & ... & 
K_{n+L_{1}-1}\left( \xi _{L_{2}},\eta _{j_{L_{2}}}^{\ast }\right)%
\end{array}%
\right\vert  \TCItag{$A2$}
\end{eqnarray}%
Concerning the indices $\left\{ j_{1},...,j_{L_{2}-L_{1}}\right\} $ we
develop $\left[ n,L\right] ^{\ast }\left( \eta _{i}\right) $ defined in (4)
in regards to the elements $\ p_{n}^{\ast }\left( \eta _{j}\right) $

\begin{equation}
\left[ n,L\right] ^{\ast }\left( \eta _{i}\right) =\sum_{j=1}^{L}\left(
-\right) ^{L-j}\ \left[ n+1,L-1\right] _{\widehat{j}}^{\ast }\left( \eta
_{i}\right) \ \ p_{n}^{\ast }\left( \eta _{j}\right)  \tag{$A3$}
\end{equation}%
where the notation $\left[ n+1,L-1\right] _{\widehat{j}}^{\ast }\left( \eta
_{i}\right) ,$ already described below (61), means that the column relative
to the variable $\eta _{j}^{\ast }$ has been ommitted in the determinant;
concerning the indices $\left\{ j_{L_{2}-L_{1}+1},...,j_{L_{2}}\right\} \ $%
we use the antisymmetry of the determinant to order the set $\left\{ \eta
_{j_{L_{2}-L_{1}+1}}^{\ast },...,\eta _{j_{L_{2}}}^{\ast }\right\} $ into
the set $\left\{ \eta _{1}^{\ast },...,\eta _{L_{1}}^{\ast }\right\} .$\ We
get%
\begin{eqnarray}
&<&n,L_{2}>\left( \xi _{i};\eta _{i}^{\ast }\right) =\left( -\right) ^{\frac{%
L_{1}\left( L_{1}-1\right) }{2}}\ \dprod\limits_{i=0}^{L_{1}-1}\frac{h_{n+i}%
}{\left[ n+i,L_{1}\right] ^{\ast }\left( \eta _{i}\right) }  \nonumber \\
&&\times \ \left\vert 
\begin{array}{ccc}
\left[ n+L_{1},L_{1}-1]_{\widehat{1}}^{\ast }\left( \eta _{i}\right) \right]
& ... & \left[ n+L_{1},L_{1}-1]_{\widehat{L_{1}}}^{\ast }\left( \eta
_{i}\right) \right] \\ 
... & ... & ... \\ 
\left[ n+1,L_{1}-1]_{\widehat{1}}^{\ast }\left( \eta _{i}\right) \right] & 
... & \left[ n+1,L_{1}-1]_{\widehat{L_{1}}}^{\ast }\left( \eta _{i}\right) %
\right]%
\end{array}%
\right\vert  \nonumber \\
&&\ \ \ \times \ \ \ \left\vert 
\begin{array}{ccc}
p_{n+L_{2}-1}\left( \xi _{1}\right) & ... & p_{n+L_{2}-1}\left( \xi
_{L_{2}}\right) \\ 
... & ... & ... \\ 
p_{n+L_{1}}\left( \xi _{1}\right) & ... & p_{n+L_{1}}\left( \xi
_{L_{2}}\right) \\ 
K_{n+L_{1}-1}\left( \xi _{1},\eta _{1}^{\ast }\right) & ... & 
K_{n+L_{1}-1}\left( \xi _{L_{2}},\eta _{1}^{\ast }\right) \\ 
... & ... & ... \\ 
K_{n+L_{1}-1}\left( \xi _{1},\eta _{L_{1}}^{\ast }\right) & ... & 
K_{n+L_{1}-1}\left( \xi _{L_{2}},\eta _{L_{1}}^{\ast }\right)%
\end{array}%
\right\vert  \TCItag{$A4$}
\end{eqnarray}%
It is the purpose of 3$%
{{}^\circ}%
)$ of part 1 of this appendix to prove that%
\begin{eqnarray}
&&\left\vert 
\begin{array}{ccc}
\left[ n+L_{1},L_{1}-1]_{\widehat{1}}^{\ast }\left( \eta _{i}\right) \right]
& ... & \left[ n+L_{1},L_{1}-1]_{\widehat{L_{1}}}^{\ast }\left( \eta
_{i}\right) \right] \\ 
... & ... & ... \\ 
\left[ n+1,L_{1}-1]_{\widehat{1}}^{\ast }\left( \eta _{i}\right) \right] & 
... & \left[ n+1,L_{1}-1]_{\widehat{L_{1}}}^{\ast }\left( \eta _{i}\right) %
\right]%
\end{array}%
\right\vert  \nonumber \\
&=&\left( -\right) ^{\frac{L_{1}\left( L_{1-1}\right) }{2}%
}\dprod\limits_{i=1}^{L_{1}-1}\left[ n+i,L_{1}\right] ^{\ast }\left( \eta
_{i}\right)  \TCItag{$A5$}
\end{eqnarray}%
Using the definition (25), we just finish the proof of (62).

\bigskip

2$%
{{}^\circ}%
)\ $We consider $L_{2}\leq L_{1}.$From (20) and (61) we write

\begin{eqnarray}
&<&n,L_{2}>\left( \xi _{i};\eta _{i}^{\ast }\right)
=\dprod\limits_{i=0}^{L_{2}-1}\frac{h_{n+i}}{\left[ n+i,L_{1}\right] ^{\ast
}\left( \eta _{i}\right) }  \nonumber \\
&&\times \sum_{\left\{ j_{k}\right\} =1}^{L_{1}}\left( -\right)
^{L_{1}L_{2}-\sum j_{k}}\dprod\limits_{k=1}^{L_{2}}\left[ n+1+L_{2}-k,L_{1}-1%
\right] _{\widehat{j_{k}}}^{\ast }\left( \eta _{i}\right)  \nonumber \\
&&\times \left\vert \ 
\begin{array}{ccc}
K_{n+L_{2}-1}\left( \xi _{1},\eta _{j_{1}}^{\ast }\right) & ... & 
K_{n+L_{2}-1}\left( \xi _{L_{2}},\eta _{j_{1}}^{\ast }\right) \\ 
... & ... & ... \\ 
K_{n+L_{2}-1}\left( \xi _{1},\eta _{j_{L_{2}}}^{\ast }\right) & ... & 
K_{n+L_{2}-1}\left( \xi _{L_{2}},\eta _{j_{L_{2}}}^{\ast }\right)%
\end{array}%
\right\vert  \TCItag{$A6$}
\end{eqnarray}%
where we sum over all indices $\left\{ j_{1},...,j_{L_{2}}\right\} $ running
from $1$ to $L_{1}.$ In (A6), the antisymmetry of the determinant in regards
to the exchange of the variables $\eta _{j_{i}}^{\ast }\ $is such that we
may reconstruct a determinant from the sum of products $\left[
n+1+L_{2}-k,L_{1}-1\right] _{\widehat{j_{k}}}^{\ast }\left( \eta _{i}\right) 
$\bigskip

\begin{eqnarray}
&<&n,L_{2}>\left( \xi _{i};\eta _{i}^{\ast }\right)
=\dprod\limits_{i=0}^{L_{2}-1}\frac{h_{n+i}}{\left[ n+i,L_{1}\right] ^{\ast
}\left( \eta _{i}\right) }\ \sum_{1\leq j_{1}<...<j_{L_{2}}\leq L_{1}}\left(
-\right) ^{L_{1}L_{2}-\sum j_{k}}  \nonumber \\
&&\times \left\vert 
\begin{array}{ccc}
\left[ n+L_{2},L_{1}-1\right] _{\widehat{j_{1}}}^{\ast }\left( \eta
_{i}\right) & ... & \left[ n+L_{2},L_{1}-1\right] _{\widehat{j_{L_{2}}}%
}^{\ast }\left( \eta _{i}\right) \\ 
... & ... & ... \\ 
\left[ n+1,L_{1}-1\right] _{\widehat{j_{1}}}^{\ast }\left( \eta _{i}\right)
& ... & \left[ n+1,L_{1}-1\right] _{\widehat{j_{L_{2}}}}^{\ast }\left( \eta
_{i}\right)%
\end{array}%
\right\vert  \nonumber \\
&&\times \ \left\vert \ 
\begin{array}{ccc}
K_{n+L_{2}-1}\left( \xi _{1},\eta _{j_{1}}^{\ast }\right) & ... & 
K_{n+L_{2}-1}\left( \xi _{L_{2}},\eta _{j_{1}}^{\ast }\right) \\ 
... & ... & ... \\ 
K_{n+L_{2}-1}\left( \xi _{1},\eta _{j_{L_{2}}}^{\ast }\right) & ... & 
K_{n+L_{2}-1}\left( \xi _{L_{2}},\eta _{j_{L_{2}}}^{\ast }\right)%
\end{array}%
\right\vert  \TCItag{$A7$}
\end{eqnarray}%
In 3$%
{{}^\circ}%
)$ of part 1 of the appendix, we prove that for\ \ \ \ $1\leq
j_{1}<...<j_{L_{2}}\leq L_{1}$

\begin{eqnarray}
&&\left\vert 
\begin{array}{ccc}
\left[ n+L_{2},L_{1}-1\right] _{\widehat{j_{1}}}^{\ast }\left( \eta
_{i}\right) & ... & \left[ n+L_{2},L_{1}-1\right] _{\widehat{j_{L_{2}}}%
}^{\ast }\left( \eta _{i}\right) \\ 
... & ... & ... \\ 
\left[ n+1,L_{1}-1\right] _{\widehat{j_{1}}}^{\ast }\left( \eta _{i}\right)
& ... & \left[ n+1,L_{1}-1\right] _{\widehat{j_{L_{2}}}}^{\ast }\left( \eta
_{i}\right)%
\end{array}%
\right\vert  \nonumber \\
&=&\left( -\right) ^{\frac{L_{2}\left( L_{2}-1\right) }{2}%
}\dprod\limits_{i=1}^{L_{2}-1}\left[ n+i,L_{1}\right] ^{\ast }\left( \eta
_{i}\right) \ \ \ \left[ n+L_{2},L_{1}-L_{2}\right] _{\left( \widehat{j_{1}}%
,...,\widehat{j_{L_{2}}}\right) }^{\ast }\left( \eta _{i}\right) \ \ \  
\TCItag{$A8$}
\end{eqnarray}%
where the notation $\left[ n+L_{2},L_{1}-L_{2}\right] _{\left( \widehat{j_{1}%
},...,\widehat{j_{L_{2}}}\right) }^{\ast }\left( \eta _{i}\right) $ means
that the columns relative to the variables $\eta _{j_{1}}^{\ast },...,\eta
_{j_{L_{2}}}^{\ast }$have been ommitted in the determinant.\ Consequently, 
\begin{eqnarray}
&<&n,L_{2}>\left( \xi _{i};\eta _{i}^{\ast }\right) =\frac{%
\dprod\limits_{i=0}^{L_{2}-1}h_{n+i}}{\left[ n,L_{1}\right] ^{\ast }\left(
\eta _{i}\right) }  \nonumber \\
&&\times \ \sum_{1\leq j_{1}<...<j_{L_{2}}\leq L_{1}}\left( -\right) ^{\chi
\left( L_{1},L_{2}\right) -\sum j_{k}}\ \ \ \left[ n+L_{2},L_{1}-L_{2}\right]
_{\left( \widehat{j_{1}},...,\widehat{j_{L_{2}}}\right) }^{\ast }\left( \eta
_{i}\right) \ \   \nonumber \\
&&\times \ \left\vert \ 
\begin{array}{ccc}
K_{n+L_{2}-1}\left( \xi _{1},\eta _{j_{1}}^{\ast }\right) & ... & 
K_{n+L_{2}-1}\left( \xi _{1},\eta _{j_{L_{2}}}^{\ast }\right) \\ 
... & ... & ... \\ 
K_{n+L_{2}-1}\left( \xi _{L_{2}},\eta _{j_{1}}^{\ast }\right) & ... & 
K_{n+L_{2}-1}\left( \xi _{L_{2}},\eta _{j_{L_{2}}}^{\ast }\right)%
\end{array}%
\right\vert  \TCItag{$A9$}
\end{eqnarray}%
where $\chi \left( L_{1},L_{2}\right) =L_{1}L_{2}-\frac{L_{2}\left(
L_{2}-1\right) }{2}.\ $From the definition (26), the above sum is nothing
but \ $D_{n+L_{2}-1}^{\ast }\left( \eta _{i};\xi _{i}^{\ast }\right) ,$
developped relatively to the bloc $K_{n+L_{2}-1}$ by choosing the columns $%
\left\{ j_{1}<...<j_{L_{2}}\right\} $ and summing over all possible
permutations. This achieve the proof of (63).

\bigskip

3$%
{{}^\circ}%
)\ $We prove the equation (A8) which is meaningful only if $L_{1}\geq L_{2}$%
. We first note that if $L_{1}=L_{2}$\ equation (A8) is simply equation (A5)
via the convention $\left[ n,0\right] _{\left( \widehat{1},...,\widehat{L_{1}%
}\right) }=1.$

We introduce $\left( L_{1}-L_{2}\right) \ $independant functions $%
N_{i}\left( z\right) \ $for $i=1,...,L_{1}-L_{2}$ and we consider the sum%
\begin{eqnarray}
J &=&\sum_{\left\{ \alpha _{k}\right\} =1}^{L_{1}}\left( -\right)
^{L_{1}^{2}-\sum \alpha _{k}}\ \ \ p_{n+L_{1}-1}\left( \eta _{\alpha
_{1}}\right) ...p_{n}\left( \eta _{\alpha _{L_{1}}}\right) \   \nonumber \\
&&\times \ \left\vert 
\begin{array}{ccc}
N_{L_{1}-L_{2}}\left( \eta _{\alpha _{1}}\right) & ... & N_{L_{1}-L_{2}}%
\left( \eta _{\alpha _{L_{1}}}\right) \\ 
... & ... & ... \\ 
N_{1}\left( \eta _{\alpha _{1}}\right) & ... & N_{1}\left( \eta _{\alpha
_{L_{1}}}\right) \\ 
\left[ n+L_{2},L_{1}-1\right] _{\widehat{\alpha _{1}}}\left( \eta _{i}\right)
& ... & \left[ n+L_{2},L_{1}-1\right] _{\widehat{\alpha _{L_{1}}}}\left(
\eta _{i}\right) \\ 
... & ... & ... \\ 
\left[ n+1,L_{1}-1\right] _{\widehat{\alpha _{1}}}\left( \eta _{i}\right) & 
... & \left[ n+1,L_{1}-1\right] _{\widehat{\alpha _{L_{1}}}}\left( \eta
_{i}\right)%
\end{array}%
\right\vert  \TCItag{$A10$}
\end{eqnarray}%
where we sum over all indices $\left\{ \alpha _{1},...,\alpha
_{L_{1}}\right\} \;$running from $1$ to $L_{1}.\ $

\bigskip

On one hand, a direct calculation shows that $J$ is the determinant%
\begin{equation}
\left\vert 
\begin{array}{cccccc}
X_{n+L_{1}-1,L_{1}-L_{2}} & ... & X_{n+L_{2},L_{1}-L_{2}} & 
X_{n+L_{2}-1,L_{1}-L_{2}} & ... & X_{n,L_{1}-L_{2}} \\ 
... & ... & ... & ... & ... & ... \\ 
X_{n+L_{1}-1,1} & ... & X_{n+L_{2},1} & X_{n+L_{2}-1,1} & ... & X_{n,1} \\ 
0 & ... & 0 & \left[ n+L_{2}-1,L_{1}\right] \left( \eta _{i}\right) & ... & ?
\\ 
... & ... & ... & ... & ... & ... \\ 
0 & ... & 0 & 0 & ... & \left[ n,L_{1}\right] \left( \eta _{i}\right)%
\end{array}%
\right\vert  \tag{$A11$}
\end{equation}%
where%
\begin{equation}
X_{n,m}=\sum_{\alpha =1}^{L_{1}}\left( -\right) ^{L_{1}-\alpha }\
p_{n}\left( \eta _{\alpha }\right) \ \ N_{m}\left( \eta _{\alpha }\right) \ 
\tag{$A12$}
\end{equation}%
The question marks ? are easy to compute but useless and the zeroes come
from the fact that we reconstruct determinants with two lines alike. The $%
L_{2}$ bottom lines of the determinant is a $\left[ L_{2}\times L_{1}-L_{2}%
\right] \ $rectangle of zeroes and a $\left[ L_{2}\times L_{2}\right] $
triangular matrix with zeroes below the diagonal and ? above. Consequently%
\begin{equation}
J=\dprod\limits_{i=0}^{L_{2}-1}\left[ n+i,L_{1}\right] \left( \eta
_{i}\right) \ \ \ \left\vert 
\begin{array}{ccc}
X_{n+L_{1}-1,L_{1}-L_{2}} & ... & X_{n+L_{1}-1,1} \\ 
... & ... & ... \\ 
X_{n+L_{2},L_{1}-L_{2}} & ... & X_{n+L_{2},1}%
\end{array}%
\right\vert  \tag{$A13$}
\end{equation}

\bigskip

On the other hand, we may use the antisymmetry in the indices $\alpha _{i}\ $%
of the determinant in (A10) to reconstruct a determinant with the $p^{\prime
}s$ which is nothing but $\left[ n,L_{1}\right] \left( \eta _{i}\right) ;$
the quantity $J\ $is now obtained from a restricted sum $\sum_{1\leq \alpha
_{1}<...<\alpha _{L_{1}}\leq L_{1}}$ and such a restriction implies $\alpha
_{i}=i.\ \ $We may write$\ J$ as 
\begin{equation}
\left( -\right) ^{\frac{L_{1}\left( L_{1-1}\right) }{2}}\ \left[ n,L_{1}%
\right] \left( \eta _{i}\right) \ \ \left\vert 
\begin{array}{ccc}
N_{L_{1}-L_{2}}\left( \eta _{1}\right) & ... & N_{L_{1}-L_{2}}\left( \eta
_{L_{1}}\right) \\ 
... & ... & ... \\ 
N_{1}\left( \eta _{1}\right) & ... & N_{1}\left( \eta _{L_{1}}\right) \\ 
\left[ n+L_{2},L_{1}-1\right] _{\widehat{1}}\left( \eta _{i}\right) & ... & 
\left[ n+L_{2},L_{1}-1\right] _{\widehat{L_{1}}}\left( \eta _{i}\right) \\ 
... & ... & ... \\ 
\left[ n+1,L_{1}-1\right] _{\widehat{1}}\left( \eta _{i}\right) & ... & 
\left[ n+1,L_{1}-1\right] _{\widehat{L_{1}}}\left( \eta _{i}\right)%
\end{array}%
\right\vert \ \ \   \tag{$A14$}
\end{equation}

\bigskip

Then, given a set $\left\{ j_{k}\right\} $ of $L_{2}$ indices such that $%
1\leq j_{1}<...<j_{L_{2}}\leq L_{1}$ , we define $\left\{ s_{p}\right\} \ \ $%
as the set of $\left( L_{1}-L_{2}\right) $ indices complementary of $\left\{
j_{k}\right\} $ in $\left\{ 1,...,L_{1}\right\} $ and such that $1\leq
s_{1}<...<s_{L_{1}-L_{2}}\leq L_{1}$ $.$ Now, we look in both expressions
(A13 and A14) of $J$ for the coefficient of $N_{L_{1}-L_{2}}\left( \eta
_{s_{1}}\right) ...N_{1}\left( \eta _{s_{L_{1}-L_{2}}}\right) $; in the
expression (A14) we obtain%
\begin{eqnarray}
&&\left( -\right) ^{\frac{L_{1}\left( L_{1}-1\right) }{2}}\left( -\right) ^{%
\frac{\left( L_{1}-L_{2}\right) \left( L_{1}-L_{2}+1\right) }{2}-\sum
s_{p}}\ \ \ \ \ \ \ \left[ n,L_{1}\right] \left( \eta _{i}\right) \ \  
\nonumber \\
&&\times \ \left\vert 
\begin{array}{ccc}
\left[ n+L_{2},L_{1}-1\right] _{\widehat{j_{1}}}\left( \eta _{i}\right) & ...
& \left[ n+L_{2},L_{1}-1\right] _{\widehat{j_{L_{2}}}}\left( \eta _{i}\right)
\\ 
... & ... & ... \\ 
\left[ n+1,L_{1}-1\right] _{\widehat{j_{1}}}\left( \eta _{i}\right) & ... & 
\left[ n+1,L_{1}-1\right] _{\widehat{j_{L_{2}}}}\left( \eta _{i}\right)%
\end{array}%
\right\vert  \TCItag{$A15$}
\end{eqnarray}%
In the expression (A13) the corresponding coefficient is found to be 
\begin{eqnarray}
&&\left( -\right) ^{L_{1}\left( L_{1}-L_{2}\right) -\sum s_{p}\ \
}\dprod\limits_{i=0}^{L_{2}-1}\left[ n+i,L_{1}\right] \left( \eta
_{i}\right) \   \nonumber \\
&&\times \ \left\vert 
\begin{array}{ccc}
p_{n+L_{1}-1}\left( \eta _{s_{1}}\right) & ... & p_{n+L_{1}-1}\left( \eta
_{s_{L_{1}-L_{2}}}\right) \\ 
... & ... & ... \\ 
p_{n+L_{2}}\left( \eta _{s_{1}}\right) & ... & p_{n+L_{2}}\left( \eta
_{s_{L_{1}-L_{2}}}\right)%
\end{array}%
\right\vert  \TCItag{$A16$}
\end{eqnarray}

\bigskip

If we compare (A15) and (A16) we obtain (A8) up to complex conjugaison.

\bigskip

\bigskip

\textsl{Part 2:\ }We calculate $D_{n}\left( \xi _{i};\eta _{i}^{\ast
}\right) \ $\ and $D_{n}^{\ast }\left( \eta _{i};\xi _{i}^{\ast }\right) $
defined in (25-26) in terms of the generalized Schur polynomials (33). From
(25) and (14) we have in the case $L_{1}\leq L_{2}$%
\begin{equation}
D_{n}\left( \xi _{i};\eta _{i}^{\ast }\right) =\sum_{\left\{ \alpha
_{k}\right\} =0}^{n}\ \dprod\limits_{i=1}^{L_{1}}\frac{p_{\alpha _{i}}^{\ast
}\left( \eta _{i}\right) }{h_{\alpha _{i}}}\ \left\vert 
\begin{array}{ccc}
p_{n+L_{2}-L_{1}}\left( \xi _{1}\right) & ... & p_{n+L_{2}-L_{1}}\left( \xi
_{L_{2}}\right) \\ 
... & ... & ... \\ 
p_{n+1}\left( \xi _{1}\right) & ... & p_{n+1}\left( \xi _{L_{2}}\right) \\ 
p_{\alpha _{1}}\left( \xi _{1}\right) & ... & p_{\alpha _{1}}\left( \xi
_{L_{2}}\right) \\ 
... & ... & ... \\ 
p_{\alpha _{L_{1}}}\left( \xi _{1}\right) & ... & p_{\alpha _{L_{1}}}\left(
\xi _{L_{2}}\right)%
\end{array}%
\right\vert  \tag{$A17$}
\end{equation}%
where we sum over $L_{1}$ indices $\alpha _{k}$ such that $0\leq \alpha
_{k}\leq n.\ $Because of the antisymmetry in the indices $\alpha _{k}$ we
can reorganize the sum into a restricted sum and we introduce a second
determinant%
\begin{eqnarray}
D_{n}\left( \xi _{i};\eta _{i}^{\ast }\right) &=&\sum_{0\leq \alpha
_{L_{1}}<...<\alpha _{1}\leq n}\ \dprod\limits_{i=1}^{L_{1}}\frac{1}{%
h_{\alpha _{i}}}\ \left\vert 
\begin{array}{ccc}
p_{\alpha _{1}}^{\ast }\left( \eta _{1}\right) & ... & p_{\alpha _{1}}^{\ast
}\left( \eta _{L_{1}}\right) \\ 
... & ... & ... \\ 
p_{\alpha _{L_{1}}}^{\ast }\left( \eta _{1}\right) & ... & p_{\alpha
_{L_{1}}}^{\ast }\left( \eta _{L_{1}}\right)%
\end{array}%
\right\vert \   \nonumber \\
&&\times \ \left\vert 
\begin{array}{ccc}
p_{n+L_{2}-L_{1}}\left( \xi _{1}\right) & ... & p_{n+L_{2}-L_{1}}\left( \xi
_{L_{2}}\right) \\ 
... & ... & ... \\ 
p_{n+1}\left( \xi _{1}\right) & ... & p_{n+1}\left( \xi _{L_{2}}\right) \\ 
p_{\alpha _{1}}\left( \xi _{1}\right) & ... & p_{\alpha _{1}}\left( \xi
_{L_{2}}\right) \\ 
... & ... & ... \\ 
p_{\alpha _{L_{1}}}\left( \xi _{1}\right) & ... & p_{\alpha _{L_{1}}}\left(
\xi _{L_{2}}\right)%
\end{array}%
\right\vert  \TCItag{$A18$}
\end{eqnarray}

Clearly, if $n<L_{1}-1,$ there is not enough room for the indices $\alpha
_{k}$ and $D_{n}\left( \xi _{i};\eta _{i}^{\ast }\right) =0;$ if $n=L_{1}-1,$
the indices are fixed to $\alpha _{k}=L_{1}-k$ so that $D_{L_{1}-1}\left(
\xi _{i};\eta _{i}^{\ast }\right) =\dprod\limits_{k=0}^{L_{1}-1}\frac{1}{%
h_{k}}\ \ \Delta ^{\ast }\left( \eta _{i}\right) \ \Delta \left( \xi
_{i}\right) \ $where $\Delta $ is the Vandermonde determinant (34). Now, for 
$n\geq L_{1}-1$ we introduce the generalized Schur polynomials defined in
(33) and we write\ for $L_{1}\leq L_{2}$%
\begin{eqnarray}
\frac{D_{n}\left( \xi _{i};\eta _{i}^{\ast }\right) }{\Delta ^{\ast }\left(
\eta _{i}\right) \ \Delta \left( \xi _{i}\right) } &=&\sum_{\lambda \in %
\left[ L_{1}\times \left( n-L_{1}+1\right) \right] }\dprod%
\limits_{k=1}^{L_{1}}\frac{1}{h_{\lambda _{k}+L_{1}-k}}  \nonumber \\
&&\times \ P_{\lambda }^{\ast }\left( \eta _{i}\right) \ P_{\left[ \left\{
\left( L_{2}-L_{1}\right) \times \left( n-L_{1}+1\right) \right\} \cup
\lambda \right] }\left( \xi _{i}\right) \ \   \TCItag{$A19$}
\end{eqnarray}%
where $\lambda $ is a Young tableau in the rectangle $\left[ L_{1}\times
\left( n-L_{1}+1\right) \right] $ with the rows of length $\lambda
_{k}=\alpha _{k}-L_{1}+k.$

\bigskip

A similar demonstration gives for $L_{1}\geq L_{2}\ $and\ $n\geq L_{2}-1$%
\begin{eqnarray}
\frac{D_{n}^{\ast }\left( \eta _{i};\xi _{i}{}^{\ast }\right) }{\Delta
^{\ast }\left( \eta _{i}\right) \ \Delta \left( \xi _{i}\right) }
&=&\sum_{\lambda \in \left[ L_{2}\times \left( n-L_{2}+1\right) \right]
}\dprod\limits_{k=1}^{L_{2}}\frac{1}{h_{\lambda _{k}+L_{2}-k}}\   \nonumber
\\
&&\times \ P_{\left[ \left\{ \left( L_{1}-L_{2}\right) \times \left(
n-L_{2}+1\right) \right\} \cup \lambda \right] }^{\ast }\left( \eta
_{i}\right) \ \ P_{\lambda }\left( \xi _{i}\right) \ \ \ \   \TCItag{$A20$}
\end{eqnarray}%
Again, $D_{n}^{\ast }\left( \eta _{i};\xi _{i}{}^{\ast }\right) =0$ if $%
n<L_{2}-1$ and $D_{L_{2}-1}^{\ast }\left( \eta _{i};\xi _{i}{}^{\ast
}\right) =\dprod\limits_{k=0}^{L_{2}-1}\frac{1}{h_{k}}\ \ \Delta ^{\ast
}\left( \eta _{i}\right) \ \Delta \left( \xi _{i}\right) .$

\bigskip

\bigskip

\textsl{Part 3:}\textit{\ \ }We prove the relation (75). Given a Young
tableau $\lambda $ in the rectangle $\left[ L\times N\right] $ and the
corresponding Young tableau $\mu =\widetilde{\lambda ^{\prime }}\ $(as
defined in Sect. 3 (36)) in the rectangle $\left[ N\times L\right] $ we
establish a relation between the lengths of the rows of $\lambda $ and the
lengths of the rows of $\mu .$ The lengths of the rows of $\lambda $ are
defined by $L$ numbers satisfying$\ N\geq \lambda _{1}\geq ...\geq \lambda
_{L}\geq 0.\ $We define the numbers $\nu _{k}\ $as the numbers of rows of $%
\lambda $ with length $k.$ We have%
\begin{eqnarray}
\sum_{k=0}^{N}\nu _{k} &=&L  \TCItag{$A21$} \\
\sum_{k=0}^{N}k\ \nu _{k} &=&\sum_{i=1}^{L}\lambda _{i}=\left\vert \lambda
\right\vert  \TCItag{$A22$}
\end{eqnarray}%
where $\left\vert \lambda \right\vert $ is the surface of the Young tableau $%
\lambda .\ $Finally, we define%
\begin{equation}
\sigma _{k}=\sum_{i=0}^{k}\nu _{i}  \tag{$A23$}
\end{equation}%
so that $\sigma _{N}=L$; the sequence $\left\{ \sigma _{k}\right\} $ is a
weakly increasing sequence as $k$ runs from $0$ to $N.\ $We have the
following properties: for $0\leq k\leq N,$ if $\nu _{k}>0,$ then%
\begin{equation}
\lambda _{L-\sigma _{k}+1}=\lambda _{L-\sigma _{k}+2}=...=\lambda _{L-\sigma
_{k-1}\ \ \ \ \ }=k  \tag{$A24$}
\end{equation}%
with $\sigma _{-1}=0$ by convention. Now, we define the numbers%
\begin{equation}
\alpha _{i}=\lambda _{i}+L-i\ \ \ \ \ \text{for\ }i=1,...,L  \tag{$A25$}
\end{equation}%
The sequence $\left\{ \alpha _{i}\right\} $ is a strictly decreasing
sequence as $i$ runs from $1$ to $L.$ Because of (A24), If\ $\nu _{k}>0$%
\[
\ \alpha _{L-\sigma _{k}+1}\ \text{to\ }\alpha _{L-\sigma _{k-1}}\text{ are
consecutive numbers from }\sigma _{k}+k-1\ \text{to }\sigma _{k-1}+k 
\]%
\begin{equation}
\tag{$A26$}
\end{equation}%
Clearly, the numbers $\alpha _{i}$ are all numbers from $N+L-1$ to $0$ at
the exception of the set $\left\{ \sigma _{k}+k\right\} \ $for $%
k=0,...,N-1.\ $We note that if$\ \nu _{k}=0$ with $\nu _{k+1}\nu _{k-1}>0,$
we get two consecutive missing numbers $\sigma _{k}+k$ and $\sigma
_{k-1}+k-1=\sigma _{k}+k-1;$ if $p$ consecutive $\nu ^{\prime }s\ $are nul,
we get $\left( p+1\right) $ consecutive numbers missing in the set $\left\{
\alpha _{i}\right\} $ and belonging to the set $\left\{ \sigma
_{k}+k\right\} .$ The set $\left\{ \sigma _{k}+k\right\} $ is a strictly
increasing sequence as $k$ runs from $0$ to $N-1.$ We may write\bigskip\ 
\begin{equation}
\left\{ \alpha _{i}\right\} =\left\{ 0,1,2,...,N+L-1\right\} -\left\{ \cup
_{k=0}^{N-1}\left( \sigma _{k}+k\right) \right\}  \tag{$A27$}
\end{equation}

\bigskip

We now consider the Young tableau $\lambda ^{\prime }$ in the rectangle $%
\left[ N\times L\right] $ and symmetric of $\lambda $ with regards to the
diagonal line. It is easy to convince oneself that%
\begin{equation}
\lambda _{k}^{\prime }=\sum_{i=k}^{N}\nu _{i}=L-\sigma _{k-1}\ \ \ \ \ \text{%
for }k=1,...,N  \tag{$A28$}
\end{equation}%
and consequently, if $\mu =\widetilde{\lambda ^{\prime }}\ $the lengths of
the rows of $\mu $ are%
\begin{equation}
\mu _{k}=L-\lambda _{N-k+1}^{\prime }=\sigma _{N-k}\ \ \ \ \ \text{for }%
k=1,...,N  \tag{$A29$}
\end{equation}

\bigskip The sequence $\left\{ \mu _{k}\right\} $ is a weakly decreasing
sequence as $k$ runs from $1$ to $N.\ $\ We define 
\begin{equation}
\beta _{k}=\mu _{k}+N-k=\sigma _{N-k}+N-k\ \ \ \ \ \text{for }k=1,...,N 
\tag{$A30$}
\end{equation}%
The sequence $\left\{ \beta _{k}\right\} $is a strictly decreasing sequence
as $k$ runs from $1$ to $N.$

\bigskip

We just proved that%
\begin{eqnarray}
\left\{ \alpha _{i}\right\} \cup \left\{ \beta _{k}\right\} &=&\left\{
0,1,2,...,N+L-1\right\}  \TCItag{$A31$} \\
\left\{ \alpha _{i}\right\} \cap \left\{ \beta _{k}\right\} &=&\phi 
\TCItag{$A32$}
\end{eqnarray}%
and this concludes the proof of (75).

\bigskip

\textsl{Part 4:\ }This part of the appendix is devoted to the proof of the
equations (93-95) and (97).

\bigskip

\bigskip 1$%
{{}^\circ}%
)\ $We first mention a generalisation of Christoffel-Darboux relation. \
Given two sets of $K$ variables $\left( \xi _{1}...\xi _{K}\right) $ and $%
\left( \eta _{1}...\eta _{K}\right) ,$ from (90-91) we may evaluate the
following integrals as%
\begin{eqnarray}
I_{N} &=&\int ...\int \dprod\limits_{i=1}^{N}d\mu \left( x_{i}\right) \
\dprod\limits_{i<j}\left( x_{i}-x_{j}\right) ^{2}\
\dprod\limits_{i=1}^{N}\dprod\limits_{a=1}^{K}\left( x_{i}-\xi _{a}\right) \
\ \dprod\limits_{i=1}^{N}\dprod\limits_{b=1}^{K}\left( x_{i}-\eta _{b}\right)
\nonumber \\
&=&N!\ \frac{\dprod\limits_{i=0}^{N-1}h_{i}}{\Delta \left( \xi _{a},\eta
_{b}\right) }\ \left[ N,2K\right] \left( \xi _{a},\eta _{b}\right) 
\TCItag{$A33$}
\end{eqnarray}%
On the other hand, we may write%
\begin{equation}
\dprod\limits_{i<j}\left( x_{i}-x_{j}\right) \
\dprod\limits_{i=1}^{N}\dprod\limits_{a=1}^{K}\left( x_{i}-\xi _{a}\right) =%
\frac{\Delta \left( x_{i},\xi _{a}\right) }{\Delta \left( \xi _{a}\right) } 
\tag{$A34$}
\end{equation}%
and similarly for the $\eta _{b}$ part, we may introduce the Vandermonde
determinants of the orthogonal polynomials $p_{n}\left( x\right) $ and
integrate; we get%
\begin{eqnarray}
I_{N} &=&N!\ \frac{\dprod\limits_{i=0}^{N+K-1}h_{i}}{\Delta \left( \xi
_{a}\right) \Delta \left( \eta _{b}\right) }\sum_{0\leq j_{1}<...<j_{K}\leq
N+K-1}\ \dprod\limits_{\alpha =1}^{K}\frac{1}{h_{j_{\alpha }}}  \nonumber \\
&&\times \ \left\vert 
\begin{array}{ccc}
p_{j_{K}}\left( \xi _{1}\right) & ... & p_{j_{K}}\left( \xi _{K}\right) \\ 
... & ... & ... \\ 
p_{j_{1}}\left( \xi _{1}\right) & ... & p_{j_{1}}\left( \xi _{K}\right)%
\end{array}%
\right\vert .\left\vert 
\begin{array}{ccc}
p_{j_{K}}\left( \eta _{1}\right) & ... & p_{j_{K}}\left( \eta _{K}\right) \\ 
... & ... & ... \\ 
p_{j_{1}}\left( \eta _{1}\right) & ... & p_{j_{1}}\left( \eta _{K}\right)%
\end{array}%
\right\vert  \TCItag{$A35$}
\end{eqnarray}%
The sum in (A35) is nothing but $\det \ K_{N+K-1}\left( \xi _{a},\eta
_{b}\right) $ where%
\begin{equation}
K_{n}\left( \xi _{a},\eta _{b}\right) =\sum_{i=0}^{n}\frac{1}{h_{i}}%
p_{i}\left( \xi _{a}\right) \ p_{i}\left( \eta _{b}\right)  \tag{$A36$}
\end{equation}

\bigskip

\bigskip We just proved that%
\begin{equation}
\frac{\Delta \left( \xi _{a}\right) \Delta \left( \eta _{b}\right) }{\Delta
\left( \xi _{a},\eta _{b}\right) }\ \left[ N,2K\right] \left( \xi _{a},\eta
_{b}\right) =\dprod\limits_{i=N}^{N+K-1}h_{i}\ \ \ \det \ K_{N+K-1}\left(
\xi _{a},\eta _{b}\right)  \tag{$A37$}
\end{equation}%
For $K=1,$ equation (A37) is nothing but Christoffel-Darboux relation

\begin{equation}
\frac{1}{\xi -\eta }\left\vert 
\begin{array}{cc}
p_{N+1}\left( \xi \right) & p_{N+1}\left( \eta \right) \\ 
p_{N}\left( \xi \right) & p_{N}\left( \eta \right)%
\end{array}%
\right\vert =h_{N}\ \ K_{N}\left( \xi ,\eta \right)  \tag{$A38$}
\end{equation}%
Also, for $N=0$ we have%
\begin{equation}
\Delta \left( \xi _{a}\right) \Delta \left( \eta _{b}\right)
=\dprod\limits_{i=0}^{K-1}h_{i}\ \ \ \det \ K_{K-1}\left( \xi _{a},\eta
_{b}\right)  \tag{$A39$}
\end{equation}

\bigskip

2$%
{{}^\circ}%
)\ $We wish to evaluate the integrals 
\begin{equation}
R_{n}^{\left( b\right) }\left( \xi _{a},y\right) =\int d\mu \left( x\right)
\ \frac{\dprod\limits_{a=1}^{L}\left( x-\xi _{a}\right) }{x-y}\ K_{n}\left(
\xi _{b},x\right) \ \ \text{for }b=1,...,L  \tag{$A40$}
\end{equation}%
Clearly,%
\begin{equation}
f_{0}\left( y,\theta \right) =\int d\mu \left( x\right) \ \frac{1}{x-y}\
K_{n}\left( \theta ,x\right) =2i\pi \ H_{n}\left( \theta ;y\right) 
\tag{$A41$}
\end{equation}%
where $H_{n}\left( \theta ,y\right) $ is defined in (94) in terms of the
Cauchy transform. Also,%
\begin{equation}
f_{1}\left( \xi ,y,\theta \right) =\int d\mu \left( x\right) \ \frac{x-\xi }{%
x-y}\ K_{n}\left( \theta ,x\right) =\left( y-\xi \right) \ 2i\pi \
H_{n}\left( \theta ;y\right) +1  \tag{$A42$}
\end{equation}%
We observe the recurrence relation%
\begin{equation}
f_{L}\left( \xi _{1},...,\xi _{L},y,\theta \right) =\left( y-\xi _{L}\right)
\ f_{L-1}\left( \xi _{1},...,\xi _{L-1},y,\theta \right) +const(\xi
_{1},...,\xi _{L-1},\theta )  \tag{$A43$}
\end{equation}%
where constant means $y$ independent. It is easy to see that the constant in
(A43) is nul if the two following conditions are fulfilled: first, $n\geq
L-1,$ second $\ \theta =\xi _{b}$ for $b=1,...,L-1$. Of course, $f_{L}\left(
\xi _{1},...,\xi _{L},y,\theta \right) $ is completely symmetric in the $\xi 
$'s and the same result is valid for $\theta =\xi _{L}\ $by simply
performing the recurrence (A43)\ from $\left( y-\xi _{b\neq L}\right) .\ $%
Consequently, 
\begin{equation}
R_{n}^{\left( b\right) }\left( \xi _{a},y\right)
=\dprod\limits_{a=1}^{L}\left( y-\xi _{a}\right) \ \ \left[ 2i\pi \
H_{n}\left( \xi _{b};y\right) +\frac{1}{y-\xi _{b}}\right] \ \ \ \text{for }%
b=1,...,L\leq n+1  \tag{$A44$}
\end{equation}%
Finally, using the same recurrence (A43) again, we may write for $%
b=1,...,L>n+1$ 
\begin{eqnarray}
R_{n}^{\left( b\right) }\left( \xi _{a},y\right)
&=&\dprod\limits_{a=1}^{L}\left( y-\xi _{a}\right) \ \ \left[ 2i\pi \
H_{n}\left( \xi _{b};y\right) +\frac{1}{y-\xi _{b}}\right] \   \nonumber \\
&&+\Pi _{L-n-2}^{\left( b\right) }\left( y,\xi _{a}\right) \ \ \ \  
\TCItag{$A45$}
\end{eqnarray}%
where $\Pi _{L-n-2}^{\left( b\right) }\left( y,\xi _{\alpha }\right) $ is a
polynomial in $y$ of degree $L-n-2.$

\bigskip

3$%
{{}^\circ}%
)\ $Now, we prove (93) for $N>M$ and $L=M.$ We introduce the Cauchy
transform (83) and we write%
\begin{equation}
\left[ N-M,2M\right] \left( \xi _{i};y_{j}\right)
=\dprod\limits_{j=1}^{M}\left( \frac{1}{2i\pi }\int \frac{d\mu \left(
x_{j}\right) }{x_{j}-y_{j}}\right) \ \ \ \ \left[ N-M,2M\right] \left( \xi
_{i},x_{j}\right)  \tag{$A46$}
\end{equation}%
From (A37), (A46) becomes%
\begin{eqnarray}
\left[ N-M,2M\right] \left( \xi _{i};y_{j}\right) &=&\left( -\right) ^{M}\
\dprod\limits_{k=N-M}^{N-1}h_{k}\ \dprod\limits_{j=1}^{M}\left( \frac{1}{%
2i\pi }\int \frac{d\mu \left( x_{j}\right) }{x_{j}-y_{j}}\dprod%
\limits_{i=1}^{M}\left( x_{j}-\xi _{i}\right) \right) \   \nonumber \\
&&\times \ \det \ K_{N-1}\left( \xi _{k},x_{j}\right)  \TCItag{$A47$}
\end{eqnarray}%
For each column of the determinant, we can perform the integration by
applying (A40) and (A44)%
\begin{eqnarray}
\left[ N-M,2M\right] \left( \xi _{i};y_{j}\right) &=&\left( \frac{-1}{2i\pi }%
\right) ^{M}\ \dprod\limits_{k=N-M}^{N-1}h_{k}\ \
\dprod\limits_{j=1}^{M}\dprod\limits_{i=1}^{M}\left( y_{j}-\xi _{i}\right) \
\   \nonumber \\
&&\times \det \ \left[ 2i\pi \ H_{N-1}\left( \xi _{k};y_{j}\right) +\frac{1}{%
y_{j}-\xi _{k}}\right]  \TCItag{$A48$}
\end{eqnarray}%
This achieve the proof of equation (93).\bigskip

\bigskip

Next, we prove (95)\ for $N\leq M.$ The determinant $\left[ N-M,L+M\right]
\left( \xi _{i};y_{j}\right) $ can be developped in regards to the variables 
$p_{q}\left( y_{j}\right) $ that is in regards to the $C_{M}^{N}$
corresponding minors. Each minor is characterized by a set of $N\ $missing
variables $\left( y_{j_{1}},...,y_{j_{N}}\right) \ $and such a minor is
multiplied in the original determinant by 
\begin{equation}
\left[ 0,L+N\right] \left( \xi _{i};y_{j_{\alpha }}\right) =\left( -\right)
^{LN}\ \Delta \left( \xi _{i}\right) \dprod\limits_{\alpha =1}^{N}\left( 
\frac{1}{2i\pi }\int \frac{d\mu \left( x_{\alpha }\right) }{x_{\alpha
}-y_{j_{\alpha }}}\dprod\limits_{i=1}^{L}\left( x_{\alpha }-\xi _{i}\right)
\right) \ \ \ \Delta \left( x_{b}\right)  \tag{$A49$}
\end{equation}%
Then, in (A49) we transform the Vandermonde $\Delta \left( x_{b}\right) $
according to (A39)%
\begin{eqnarray}
&&\left[ 0,L+N\right] \left( \xi _{i};y_{j_{\alpha }}\right)  \nonumber \\
&=&\left( -\right) ^{LN}\ \frac{\Delta \left( \xi _{i}\right) }{\Delta
\left( \theta _{a}\right) }\dprod\limits_{k=0}^{N-1}h_{k}\ \
\dprod\limits_{\alpha =1}^{N}\left( \frac{1}{2i\pi }\int \frac{d\mu \left(
x_{\alpha }\right) }{x_{\alpha }-y_{j_{\alpha }}}\dprod\limits_{i=1}^{L}%
\left( x_{\alpha }-\xi _{i}\right) \right) \   \nonumber \\
&&\times \det \ K_{N-1}\left( \theta _{a},x_{b}\right)  \TCItag{$A50$}
\end{eqnarray}%
where we have introduced $N$ variables $\theta _{a}$ at wishes. However, if
we choose the set $\left\{ \theta _{a}\right\} \subset \left( \xi
_{1},...,\xi _{L}\right) $ we may apply (A40) and (A45) in order to perform
the integrations%
\begin{equation}
\left[ 0,L+N\right] \left( \xi _{i};y_{j_{\alpha }}\right) =\frac{\left(
-\right) ^{LN}}{\left( 2i\pi \right) ^{N}}\ \frac{\Delta \left( \xi
_{i}\right) }{\Delta \left( \theta _{a}\right) }\dprod%
\limits_{k=0}^{N-1}h_{k}\ \det \ U_{N-1}\left( \theta _{a},y_{j_{\alpha
}}\right)  \tag{$A51$}
\end{equation}%
where%
\begin{eqnarray}
U_{N-1}\left( \theta _{a},y_{j_{\alpha }}\right)
&=&\dprod\limits_{i=1}^{L}\left( y_{j_{\alpha }}-\xi _{i}\right) \ \ \left[
2i\pi \ H_{N-1}\left( \theta _{a};y_{j_{\alpha }}\right) +\frac{1}{%
y_{j_{\alpha }}-\theta _{a}}\right] \   \nonumber \\
&&+\Pi _{L-N-1}^{\left( a\right) }\left( y_{j_{\alpha }},\xi _{k}\right) 
\TCItag{$A52$}
\end{eqnarray}%
and where the polynomial $\Pi _{L-N-1}^{\left( a\right) }\left( y_{j_{\alpha
}},\xi _{k}\right) $ is of degree $\left( L-N-1\right) $ in $y_{j_{\alpha }}$
and is nul if $L<N-1.\ $Now, if we replace $\left[ 0,L+N\right] \left( \xi
_{i};y_{j_{\alpha }}\right) $ by (A51)\ in the original determinant
correspondingly to each minor we obtain%
\begin{eqnarray}
&&\left[ N-M,L+M\right] \left( \xi _{i};y_{j}\right)  \nonumber \\
&=&\frac{\left( -\right) ^{LN}}{\left( 2i\pi \right) ^{N}}\ \frac{\Delta
\left( \xi _{i}\right) }{\Delta \left( \theta _{a}\right) }%
\dprod\limits_{k=0}^{N-1}h_{k}\ \left\vert 
\begin{array}{ccc}
U_{N-1}\left( \theta _{1},y_{1}\right) & ... & U_{N-1}\left( \theta
_{1},y_{M}\right) \\ 
... & ... & ... \\ 
U_{N-1}\left( \theta _{N},y_{1}\right) & ... & U_{N-1}\left( \theta
_{N},y_{M}\right) \\ 
p_{M-N-1}\left( y_{1}\right) & ... & p_{M-N-1}\left( y_{M}\right) \\ 
... & ... & ... \\ 
p_{0}\left( y_{1}\right) & ... & p_{0}\left( y_{M}\right)%
\end{array}%
\right\vert  \TCItag{$A53$}
\end{eqnarray}%
If $M=N\ \ $the part $p_{q}\left( y_{j}\right) $ is absent from the
determinant (A53). We note that if $L\leq M,$ the polynomial $\ \Pi
_{L-N-1}^{\left( a\right) }\left( y_{j},\xi _{k}\right) $ in $U_{N-1}\left(
\theta _{a},y_{j}\right) $ vanish in the determinant by linear combination
of the $p_{q}\left( y_{j}\right) $ for $q=0,...,L-N-1.$This achieve the
proof of (95).\bigskip

\bigskip

4$%
{{}^\circ}%
)\ $In the case $N\geq M,$ we calculate the expression $J_{N}$ in (97) by
successive derivations $\left( -\frac{\partial }{\partial \xi _{i}}\right)
_{y_{i}=\xi _{i}}\ \ $of the relation (93).

We must calculate%
\begin{equation}
\left[ \dprod\limits_{i=1}^{M}\left( -\frac{\partial }{\partial \xi _{i}}%
\right) \ A\left( y_{b},\xi _{a}\right) \ \det \left\{ \left( y_{a}-\xi
_{a}\right) \left[ 2i\pi \ H_{N-1}\left( \xi _{a};y_{b}\right) +\frac{1}{%
y_{b}-\xi _{a}}\right] \right\} \right] _{y_{k}=\xi _{k}}  \tag{$A54$}
\end{equation}%
where%
\begin{equation}
A\left( y_{b},\xi _{a}\right) =\frac{\dprod\limits_{b=1}^{M}\dprod\limits_{a%
\neq b}\left( y_{b}-\xi _{a}\right) }{\Delta \left( \xi _{a}\right) \ \Delta
\left( y_{b}\right) }  \tag{$A55$}
\end{equation}%
In (A54), we distribute the derivatives on $A\left( y_{b},\xi _{a}\right) $
and on the determinant; the derivative $\left( -\frac{\partial }{\partial
\xi _{i}}\right) $ acts only on the line $i\ $of the determinant$.$\ After
derivation, the condition $\ y_{k}=\xi _{k}\ $\ transforms that line into%
\begin{eqnarray}
2i\pi H_{N-1}\left( \xi _{i};\xi _{j}\right) +\frac{1}{\xi _{j}-\xi _{i}}\ \
\ \ \ \ \text{for}\ j &\neq &i  \TCItag{$A56$} \\
2i\pi H_{N-1}\left( \xi _{i};\xi _{i}\right) \ \ \text{for\ }j &=&i 
\TCItag{$A57$}
\end{eqnarray}%
Now, the lines of the determinant without derivatives are $\delta _{ij}$
when $y_{k}=\xi _{k}.$ Consequently, (A54) may be written%
\begin{equation}
\sum_{I}\left[ \dprod\limits_{i\notin I}\left( -\frac{\partial }{\partial
\xi _{i}}\right) \ A\left( y_{b},\xi _{a}\right) \right] _{y_{k}=\xi _{k}}\
\det \left( I\right)  \tag{$A58$}
\end{equation}%
where $I\subset \left( \xi _{1},...,\xi _{M}\right) $ and $\det \left(
I\right) $ is a subdeterminant (A56-A57) with $i,j\in I.\ $Finally, for $%
J\subset \left( \xi _{1},...,\xi _{M}\right) $ we find%
\begin{equation}
\left[ \dprod\limits_{i\in J}\left( -\frac{\partial }{\partial \xi _{i}}%
\right) \ A\left( y_{b},\xi _{a}\right) \right] _{y_{k}=\xi _{k}}=0\ \ \ \ 
\text{if}\ card\left( J\right) \ \text{is odd}  \tag{$A59$}
\end{equation}%
and if \ $card\left( J\right) \ $is even%
\begin{equation}
\left[ \dprod\limits_{i\in J}\left( -\frac{\partial }{\partial \xi _{i}}%
\right) \ A\left( y_{b},\xi _{a}\right) \right] _{y_{k}=\xi _{k}}=\left(
-\right) ^{\frac{M\left( M-1\right) }{2}}\ \sum_{\text{all pairings in }J}\
\dprod\limits_{\left( i,j\right) \in J}\left( \frac{-1}{\left( \xi _{j}-\xi
_{i}\right) ^{2}}\right)  \tag{$A60$}
\end{equation}%
The equations (A58-A59-A60) prove the result (97).

\bigskip

\bigskip

\textbf{References}

\bigskip

\ 1. Dijkgraaf R. and Vafa C.: Nucl.\ Phys. \textbf{B644}, 3-39 (2002)

\ 2. Verbaarschot J.J.M. and Wettig T.: Ann. Rev.\ Nucl.\ Part.\ Sci.

\ \ \ \ \ \textbf{50,\ }343-410 (2000)

\ 3. Baik J., Deift P. and Strahov E.: J.\ Math.\ Phys. \textbf{44},
3657-3670 (2003)

\ 4. Akemann G. and Vernizzi G.: Nucl. Phys.\ \textbf{B660}, 532-556 (2003)

\ 5. Berenstein D., Maldacena J. and Nastase H.: JHEP\ \textbf{0204},013
(2002) \ \ \ \ \ \ \ 

\ 6. Eynard B. and Kristjansen C.: JHEP \textbf{0210}, 027 (2002)

\ 7. Szeg\"{o} G.: orthogonal polynomials, Am. Math. Soc., Colloquium

\ \ \ \ \ publications \textbf{23}, Providence (1975)

\ 8. Ginibre J.: J.\ Math.\ Phys. \textbf{6}, 440-449 (1965)

\ 9. Brezin E. and Hikami S.: Commun. Math. Phys. \textbf{214}, 111-135
(2000)

10. Uvarov V.\ B.: USSR Comput. Math. Phys. \textbf{9}, n$%
{{}^\circ}%
)$ 6, 25-36 (1969)

11. Fyodorov Y. and Strahov E.: J. of Phys. \textbf{A36}, 3203-3213 (2003) \
\ \ 

12. Dyson F.J.: J.Math.Phys. \textbf{3}, 140-175 (1962)

13. Mehta M.L.: Random matrices, Academic Press, San Diego CA, USA

\ \ \ \ \ \ (1991)

\ \ \ \ \ 

\end{document}